\documentclass[fleqn,usenatbib]{mnras}
\usepackage{newtxtext,newtxmath}
\usepackage[T1]{fontenc}
\usepackage{ae,aecompl}

\usepackage{graphicx,caption}
\usepackage{amsmath}	    % Advanced maths commands
  
\graphicspath{{./plots/}}

\newcommand\Rq{\emph{R-quantity}~}

\title[Exotic image formation in cluster lenses -- IV]
{Exotic Image Formation in Strong Gravitational Lensing by Clusters of Galaxies -- IV. 
Elliptical NFW Lenses and Hyperbolic Umbilics}

\author[A. K. Meena \& J. S. Bagla]{
Ashish Kumar Meena$^{1}$\thanks{E-mail: \href{mailto:ashishmeena766@gmail.com}{ashishmeena766@gmail.com}}
and Jasjeet Singh Bagla$^{2}$
\\
\\
$^{1}$Physics Department, Ben-Gurion University of the Negev, P.O. Box 653,
Be'er-Sheva 84105, Israel 
\\
$^{2}$Department of Physical Sciences, Indian Institute of Science Education and 
Research Mohali, Knowledge City, Sector 81, SAS Nagar, Punjab 140306, India
}

\date{Accepted XXX; Received YYY; in original form ZZZ}

\pubyear{2023}

\begin{document}
\label{firstpage}
\pagerange{\pageref{firstpage}--\pageref{lastpage}}
\maketitle

%%%%%%%%%%%%%%%%%%%%%%%%%%%%%%%%%%%%%%%%%%%%%%%%%%%%%%%%%%%%%%%%%%%%%%%%%%%%%%%%%%%%%%%%%%%
\begin{abstract}

A source lying near a hyperbolic umbilic~(HU) singularity leads to a ring-like image formation, constituting four images with high magnification factors and lying in a small region of the lens plane. Since (based on our earlier work) the observed number of HU image formations in cluster lenses is expected to increase in the future, it is timely to investigate them in more detail. Like fold and cusp singularities, HU also satisfies the magnification relation, i.e., the signed magnification sum of the four images equals zero. This work presents a detailed study of the HU magnification relation~($R_{\rm hu}$) considering the elliptical Navarro–Frenk–White~(eNFW) lens profile suitable for cluster scale dark matter halos. Our results show that for an isolated eNFW lens,~$R_{\rm hu}$ is more sensitive to ellipticity than its mass or concentration parameter. An ellipticity greater than 0.3 results in~$R_{\rm hu}$ lying close to zero with a small scatter around it. A substructure near the HU image formation causes the average~$R_{\rm hu}$ value to deviate from zero and increases the scatter, with the amount of deviation depending on the image type near which the substructure lies. However, a population of substructures in the lens plane (equivalent to the galaxies inside the cluster) does not significantly shift the average~$R_{\rm hu}$ value from zero but increases the scatter around it. We find that $R_{\rm hu} \simeq 0$ for HU image formation in the Abell~1703 cluster. Repeating this test in other clusters with HU formations can be a useful indicator of substructure in cluster halos.

\end{abstract}

%%%%%%%%%%%%%%%%%%%%%%%%%%%%%%%%%%%%%%%%%%%%%%%%%%%%%%%%%%%%%%%%%%%%%%%%%%%%%%%%%%%%%%%%%%%

%%%%%%%%%%%%%%%%%%%%%%%%%%%%%%%%%%%%%%%%%%%%%%%%%%%%%%%%%%%%%%%%%%%%%%%%%%%%%%%%%%%%%%%%%%%
\section{Introduction}
\label{sec:Intro}

Strong gravitational lensing by galaxy clusters has become an excellent probe for studying the various aspects of the Universe. 
The multiple image formation of background sources allows us to map the visible as well as dark matter in the cluster lens with great detail~\citep[e.g.,][]{
2014MNRAS.443.1549J,2016ApJ...819..114K,2017A&A...607A..93C,2022arXiv220709416B}. 
The formation of lensed images near the cluster 
centre~\citep[e.g.,][]{2018MNRAS.477..669M, 2022MNRAS.510...54A} and lensing by the cluster substructures~\citep[e.g.,][]{2020Sci...369.1347M, 2021MNRAS.505.1458B, 
2021MNRAS.504L...7R, 2021PhRvD.104j3031Y} is likely to be a  probe to determine the nature of the dark matter. 
In addition, the high magnification provided by the cluster lens allows us to 
observe distant sources that otherwise would have remained 
unobserved~\citep[e.g.,][]{2012ApJ...747....3B, 2013ApJ...762...32C, 2018ApJ...864L..22S}. 

The number of images of a strongly lensed source and their magnifications depends 
on the overall geometry of the lens system~\citep[e.g.,][]{1992grle.book.....S}.
For example, strong lensing by a single galaxy lens (assuming a non-singular lens
mass model) can lead to the formation of 
three or five images of the background source, although we mostly observe two or 
four as the central image is hard to detect~\citep[e.g.,][]{2006ApJ...638..703B, 
2012ApJ...744...41B}.
On the other hand, due to its complex mass distribution, a cluster lens can give 
rise to the formation of very complicated image 
geometries~\citep[e.g.,][]{2001ApJ...557..594R, 2009MNRAS.399....2O}.
The magnification (and also geometry) of these images depends on the distance of
the source from the caustics in the source plane: the closer the caustic, the higher 
the magnification.
A source lying near and inside a fold (cusp) caustic gives rise to a pair of 
two (three) images known as doublet (triplet) near the corresponding critical 
curve forming an arc-like structure.
It is well known that, in an ideal scenario, the signed sum of the magnification of
images belonging to doublet or triplet is always zero, 
($\Sigma_i\mu_i=0$ where $i=1,2$ for doublet and $i=1,2,3$ for 
triplet;~\citealt{1986ApJ...310..568B, 1992A&A...260....1S, 1995A&A...293....1Z, 
2001stgl.book.....P}).
However, in real gravitational lenses, the above identity is only satisfied
approximately and the deviation from zero depends on the distance of the 
source from the fold/cusp critical point and the properties of the lens potential.
Besides, a significant deviation from a zero value can be a possible signature of 
substructure near the doublet/triplet~\citep[e.g.,][]{1998MNRAS.295..587M, 
2003ApJ...598..138K, 2005ApJ...635...35K} further helping us in understanding the
nature of the dark matter~\citep[e.g.,][]{2002ApJ...572...25D, 2004ApJ...610...69K,
2012MNRAS.419..936F, 2013ApJ...773...35M, 2022arXiv221103866G}.

Focussing mainly on the quad galaxy lenses,~\citet{2003ApJ...598..138K} discussed 
in detail, the deviation of the magnification relation from zero for a source lying near 
the cusp point while varying the radial profile, ellipticity, external shear, and 
multipole density fluctuations and studied their implications in observed quad lenses.
\citet{2003ApJ...598..138K} found that cusp magnification relation is insensitive
to the radial profile of the lens, whereas it is quite sensitive to the other 
mentioned properties.
Later~\citet{2005ApJ...635...35K} performed a similar study for the fold magnification
relation.
One interesting result that~\citet{2005ApJ...635...35K} found is that the deviation 
of the fold magnification relation from zero, in addition to source distance from the 
fold caustic is also sensitive to the location of the source along the fold caustic.
In addition to magnification relation, \citet{2009ApJ...699.1720K} and \citet{2010ApJ...709..552C} 
showed that anomalies in the observed time delay measurements for close doublet and triplet images
can also hint towards the presence of substructures.
Later, \citet{2016MNRAS.461.4466C} investigated the observed flux ratio anomalies
in \textit{quad} image formations finding that the observed quad lens system flux
ratios do not agree very well with the ones predicted using quad lens models.
However, such a discrepancy can arise due to various reasons~(as discussed in the
same work) and needs further 
analysis to determine the exact cause of the observed discrepancy.

Unlike fold and cusp singularities, which are always present in strong lensing, higher-order
catastrophes (also known as point singularities) like the swallowtail, hyperbolic 
umbilic~(hereafter HU), and elliptic umbilic only appear for specific lens system 
geometries and remain very sensitive to the lens system parameters.
As shown in~\citet{2020MNRAS.492.3294M}, a very straightforward way to locate all
the point singularities (at all possible source redshifts) that a lens has to offer
is to construct a \emph{singularity map} in the image plane.
In addition to the point singularities, a singularity map also locates all the cusp
points in the image plane that a lens has to offer.
As discussed in~\citet{2021MNRAS.503.2097M} and \citet{2021MNRAS.506.1526M}, the 
number of image formations near these point singularities in the cluster lenses is 
expected to be large.
Until recently, only one HU image formation had been documented in the 
literature~\citep[e.g.,][]{2008A&A...489...23L}. 
However, a recent study by \citet{2023arXiv230309568L} has discovered three 
additional HU image formations within a single cluster. 
The study also reported the identification of 10 more HU systems, which are 
expected to be the subject of their upcoming research.
These point singularities come with a characteristic image formation with all 
images (which are part of the characteristic image formation) lying in a relatively 
small region of the lens plane (compared to the generic five image formations) 
leading to smaller relative time delay values and high magnification 
factors~\citep{2022MNRAS.515.4151M}.
In addition to the above-mentioned properties, these characteristic image formations
corresponding to these point singularities also satisfy the magnification relation 
($\Sigma_i\mu_i=0$~where $i=1,2,3,4$) similar to fold and cusp
as shown in~\citet{2009JMP....50c2501A}.
The preliminary analysis done in~\citet{2009JMP....50c2501A} finds that the magnification 
relation near HU is more stable (smaller deviation from zero) compared to fold and 
cusp and applicable even when the source is relatively far from the caustics.

Considering the facts that HU magnification relation is more stable and
we expect to detect more such systems in the future, an interesting question arises: 
\emph{can we use HU image formations to detect the substructures in the lens?}
The question becomes even more appealing as the HU characteristic image formation consists
four images near which we can look for substructures, unlike doublet/triplet
where we only have two/three images.
\citet{2009JMP....50c2501A} briefly studied the behaviour of flux ratio and the 
magnification relation in HU configuration considering a simple lens mass model of a 
singular isothermal ellipsoid (SIE) primarily focusing on the galaxy 
scale lenses.
In principle, the above lens model does not give rise to the HU image formation but
HU magnification relation can be applied to similar quad image formation in the galaxy
scale lenses.
Recent work by \citet{2023arXiv230309568L} contains a preliminary analysis in the 
same direction where the authors utilized the newly identified HU image formation and 
try to put lower bound on the mass of the substructure that can be detected by examining 
astrometric and flux ratio anomalies.
However, keeping in mind the extreme sensitivity of point singularities to the lens 
parameter and scarcity of earlier works in literature, further study is required 
to calibrate the use of point singularities to probe substructures in the 
galaxy~(or galaxy-cluster) scale lenses.

In our current work, we investigate the possibility of using HU image formation to
detect the substructures in the lens by focusing on the corresponding magnification 
relation.
So far, all the HU image formations are detected in the cluster lenses; hence, we 
only consider lensing by cluster scale halo modelled using 
Navarro–Frenk–White~\citep[NFW;][]{1996ApJ...462..563N, 1997ApJ...490..493N, 
2000ApJ...534...34W} profile.
We start by revisiting the characteristic image formation near HU singularities for an
elliptical NFW~(eNFW) lens and discuss the effect of variation in the source position 
and source redshift on the HU image formation. 
Since the redshift at which HU gets critical~($z_{\rm hu}$) depends on different lens 
parameters, we have also investigated the dependency of $z_{\rm hu}$ on various
eNFW lens parameters.
Such an analysis helps us in deducing the range of various lens parameters preferable 
to give rise to an HU image formation.
To determine the usefulness of the HU image formation in detecting substructures in the
lens, we study the HU magnification relation for an isolated eNFW lens, eNFW lens
in the presence of external shear and substructure(s). 
We note that, in principle, the flux-ratio anomalies can arise from various sources, 
like, microlensing due to stars, milli-lensing by dark matter halos, or large non-zero 
multi-pole density fluctuations.
All of these sources can be together considered as ``substructures'' and observation
of an anomalous lens system does not allow us to discriminate them without additional 
information~\citep[e.g.,][]{2003ApJ...598..138K, 2005ApJ...635...35K}.
In this work, we specifically focus on dark matter halos as substructures and omit
other possible sources which can lead to flux ratio anomalies.
Finally, we estimate the~$R_{\rm hu}$ value for HU image formation in Abell~1703
galaxy cluster.
In addition, we also present three new candidates for the HU image formation identified
in two different galaxy clusters.

This paper is organized as follows:
In Section~\ref{sec:Basics}, we briefly review the relevant basics of gravitational 
lensing.
Section~\ref{sec:ImageHU} discusses the characteristic image formation near HU and its
variation with source position and source redshift.
In Section~\ref{sec:OneNFW}, we study various HU properties as a function of a single
eNFW lens parameters.
Subsection~\ref{ssec:critz_enfw} discusses the critical HU redshift as a function of 
different eNFW lens parameters.
In Subsection~\ref{ssec:rhu_enfw_eps},~\ref{ssec:rhu_enfw_mcvir},~\ref{ssec:rhu_enfw_ee}, 
we study the HU magnification relation as a function of ellipticity, mass/concentration
parameter, and external effects for a single eNFW lens.
In Subsection~\ref{sec:OneNFWSub}, we study the effect of one substructure on the HU 
magnification relation for a single eNFW lens.
Section~\ref{sec:OneNFWMultipleSub} discusses the HU magnification relation in the 
presence of multiple galaxy-scale substructures randomly distributed in the lens plane.
In Section~\ref{sec:a1703}, we estimate the~$R_{\rm hu}$ for the HU image formation in
Abell~1703 galaxy cluster.
In Section~\ref{sec:NewHU}, we present three new HU image formation candidates
identified in two different clusters.
Conclusions, along with the possible future directions, are presented in Section~\ref{sec:conclusions}.
The cosmological parameters used in this work to estimate the various quantities are:
$H_0=70\:{\rm km\:s^{-1}\:Mpc^{-1}}$, $\Omega_{m}=0.3$, and $\Omega_{\Lambda}=0.7$.
The lens redshift is fixed to~$z_l=0.4$, and the lens and source plane resolution is 
set to $0\farcs03$ unless mentioned otherwise.

%%%%%%%%%%%%%%%%%%%%%%%%%%%%%%%%%%%%%%%%%%%%%%%%%%%%%%%%%%%%%%%%%%%%%%%%%%%%%%%%%%%%%%%%%%%
\section{Basics of Lensing}
\label{sec:Basics}

In gravitational lensing, the lens equation represents a mapping between the lens and
source plane~\citep[e.g.,][]{1992grle.book.....S}.
In angular coordinates, the lens equation can be written as 
\begin{equation}
    \pmb{\beta}=\pmb{\Theta}-\frac{D_{ds}}{D_s}\nabla\Psi(\pmb{\Theta}),
    \label{eq:LensEq}
\end{equation}
where $\pmb{\beta}$ and $\pmb{\Theta}$ are two dimensional angular vectors in the
source and lens plane, respectively.
$\nabla\Psi(\pmb{\Theta})$ is the gradient of the scaled lens potential at 
$\pmb{\Theta}$ also known as the scaled deflection angle.
$D_{ds}$ and ${D_s}$ are angular diameter distances between lens and source and
observer and source, respectively.
Various properties of the lens equation can be described by the corresponding Jacobian,
\begin{equation}
    \begin{split}
        \mathbb{A} = \frac{\partial\pmb{\beta}}{\partial\pmb{\Theta}}
        & = \delta_{ij} - \frac{D_{ds}}{D_s} \Psi_{ij} \\
        = & \begin{pmatrix}
                1 & 0 \\
                0 & 1  
            \end{pmatrix}
            - \frac{D_{ds}}{D_s}
            \begin{pmatrix}
                \kappa + \gamma_1 & \gamma_2 \\
                \gamma_2 &  \kappa - \gamma_1  
            \end{pmatrix},
    \end{split}
    \label{eq:Jacobian}
\end{equation}
where $\Psi_{ij}$ is known as the \emph{deformation tensor}.
$\kappa$ and $(\gamma_1, \gamma_2)$ are known as \emph{convergence} and \emph{shear} 
components, respectively.
Convergence describes the isotropic distortion in the lensed image whereas shear,
$\gamma \equiv (\gamma_1,\gamma_2)$, determines the anisotropic distortion in the
lensed image.
The magnification of a lensed image corresponding to a point source formed at
$\pmb{\Theta}$ in the lens plane is given as 
\begin{equation}
    \mu(\pmb{\Theta}) = \frac{1}{(1-a\alpha)(1-a\beta)},
    \label{eq:PointMag}
\end{equation}
where $\alpha$ and $\beta$ are the eigenvalues of the deformation tensor
(assuming $\alpha \geq \beta$) and $a(\equiv D_{ds}/D_s)$ is the distance ratio.
From Equation~\eqref{eq:PointMag}, we can see that when $\alpha=1/a$ or $\beta=1/a$ 
or $\alpha=1/a=\beta$, the point source magnification goes to infinity.
These points with infinite magnification form smooth closed curves in the image 
plane known as \textit{critical curves} and the corresponding curves in the 
source plane~(mapped using the lens equation; Equation~\ref{eq:LensEq}) are 
known as \textit{caustics}. 
Caustics are also closed curves but not necessarily smooth. 
In general, caustics are made of smooth segments~(folds) connected at the cusp points. 
Caustics (and the corresponding critical curves) can be further divided in two
types: radial and tangential.
A source lying near a radial (tangential) caustic leads to the formation of radially
(tangentially) elongated arc with respect to the lens centre near the corresponding
critical curve.
One can visually identify these tangential or radial arcs in simple lenses.
However, for complex cluster lenses, one needs to rely on detailed lens models.
The formation of radial arcs in cluster lenses allows us to constrain the radial 
density profile of the cluster lenses as well as gain insight into the properties
of the dark matter~\citep[e.g.,][]{2001MNRAS.325..435M, 2001ApJ...559..544M, 
2013SSRv..177...31M}.
On the other hand, the formation of tangential arcs permits us to estimate the 
total enclosed mass (at the arc position).

A source lying near and inside a fold (cusp) caustic gives rise to a pair of 
two (three) images near the corresponding critical curve.
If the source lies near a fold caustic then both of the images are, in principle, 
equally magnified and the signed sum of magnification for these two images 
vanishes, i.e., $\mu_1 + \mu_2 = 0$~\citep[e.g.,][]{2005ApJ...635...35K}.
Similarly, for a source near a cusp point, the two images forming outside the 
critical curve are equally magnified but the third image inside the critical curve 
has twice the magnification, again summing up to zero,
i.e.,~$\mu_1 + \mu_2 + \mu_3 = 0$~\citep[e.g.,][]{2003ApJ...598..138K}.
A small deviation from zero value in the observed lens systems is expected as the 
source does not necessarily lie on the caustic (but close enough to the caustic)
or we have an extended source in which case only a part of the source will lie
on the caustic. 

In strong lensing, folds and cusps are always present in the source plane; hence
known as stable singularities.
However, unstable (point) singularities like swallowtails and umbilics only occur 
for specific source redshifts.
At these point singularities, we observe emergence, destruction, or exchange of
cusps between radial and tangential caustics in the source plane.
At the swallowtail singularity, two extra cusps appear in the source plane.
On the other hand, at umbilics, we notice an exchange of cusp(s) between radial and
tangential caustics: exchange of one (three) cusp(s) at hyperbolic (elliptic)
umbilic.
We refer reader to \citet{2020MNRAS.492.3294M, 2021MNRAS.503.2097M, 2021MNRAS.506.1526M}
and \citet{2022MNRAS.515.4151M} for more detailed discussion about various 
properties of these point singularities, their appearance in cluster lens models along
with their sensitivity to lens mass reconstruction techniques and the concept of 
\textit{singularity map}.
Near these point singularities, the lens mapping can be written in terms of higher 
order derivatives of the lens potential considering the source redshift as one of the 
parameters~(see section~6.3 in~\citealt{1992grle.book.....S}).
Doing so can help us construct analytical relations for magnification and time
delays as a function of distance from the point singularity, further helping us
in understanding the properties of these point singularities.
We leave such an analytical study for future work.

%%%%%%%%%%%%%%%%%%%%%%%%%%%%%%%%%%%%%%%%%%%%%%%%%%%%%%%%%%%%%%%%%%%%%%%%%%%%%%%%%%%%%%%%%%%
\begin{figure*}
    \centering
    \includegraphics[width=13.3cm,height=20.0cm]{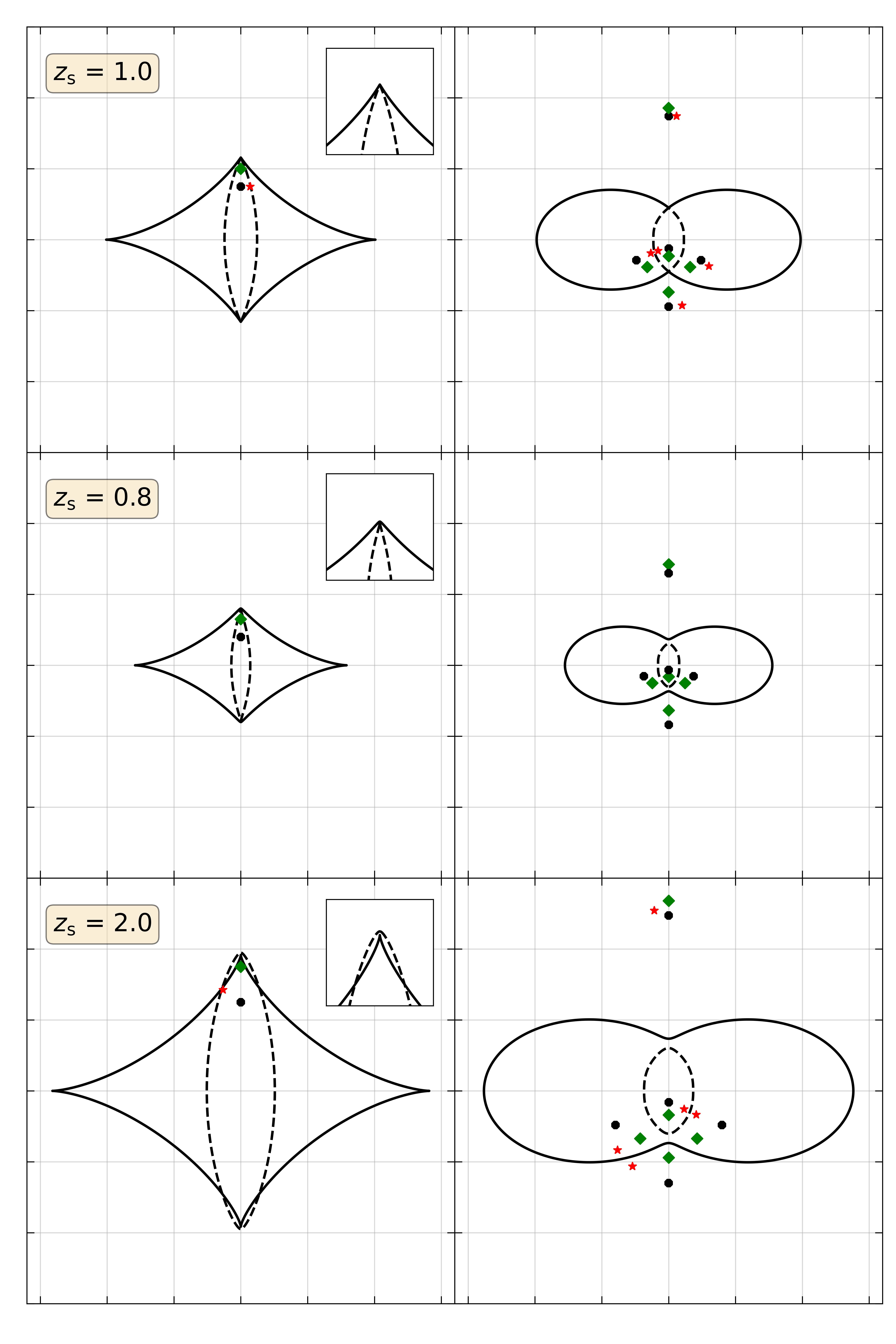}
    \caption{Examples of image formations near the HU point singularity in an isolated eNFW 
    lens. The left and right columns represent the source and image, respectively. 
    The solid and dashed black curves in the left~(right) 
    panel mark the tangential and radial caustics (critical curves), respectively. 
    In the top row, the source redshift is equal to the critical HU 
    redshift~($z_s=z_{\rm hu}$), i.e., the redshift at which HU singularity gets critical. 
    Hence, we note that in the source plane along the y-axis, the radial and 
    tangential caustics meet with each other at a cusp 
    point as can be seen in the inset plot~(also see~\citealt{2020MNRAS.492.3294M}). 
    The black dot, green diamond, and red star in the left panel represent the source 
    positions near HU singularity, and the 
    corresponding image formations are shown in the right panel. In the middle row, 
    the source redshift is smaller than the critical HU redshift~($z_s<z_{\rm hu}$) 
    and the black dot and green diamond show the source and corresponding image 
    positions in the left and right panels, respectively. In the bottom row, the 
    source redshift is greater than the critical HU redshift~($z_s>z_{\rm hu}$), and 
    the black dot, red star and green diamond show the source and corresponding 
    image positions in the left and right panels, respectively. The 
    inset plots in the left column zoon-in on the caustic structure near HU point
    singularity and combined together show the exchange of a cusp between radial 
    and tangential caustics.}
    \label{fig:ImageHU}
\end{figure*}
%%%%%%%%%%%%%%%%%%%%%%%%%%%%%%%%%%%%%%%%%%%%%%%%%%%%%%%%%%%%%%%%%%%%%%%%%%%%%%%%%%%%%%%%%%%

%%%%%%%%%%%%%%%%%%%%%%%%%%%%%%%%%%%%%%%%%%%%%%%%%%%%%%%%%%%%%%%%%%%%%%%%%%%%%%%%%%%%%%%%%%%
\begin{figure}
    \centering
    \includegraphics[width=8.5cm,height=7.0cm]{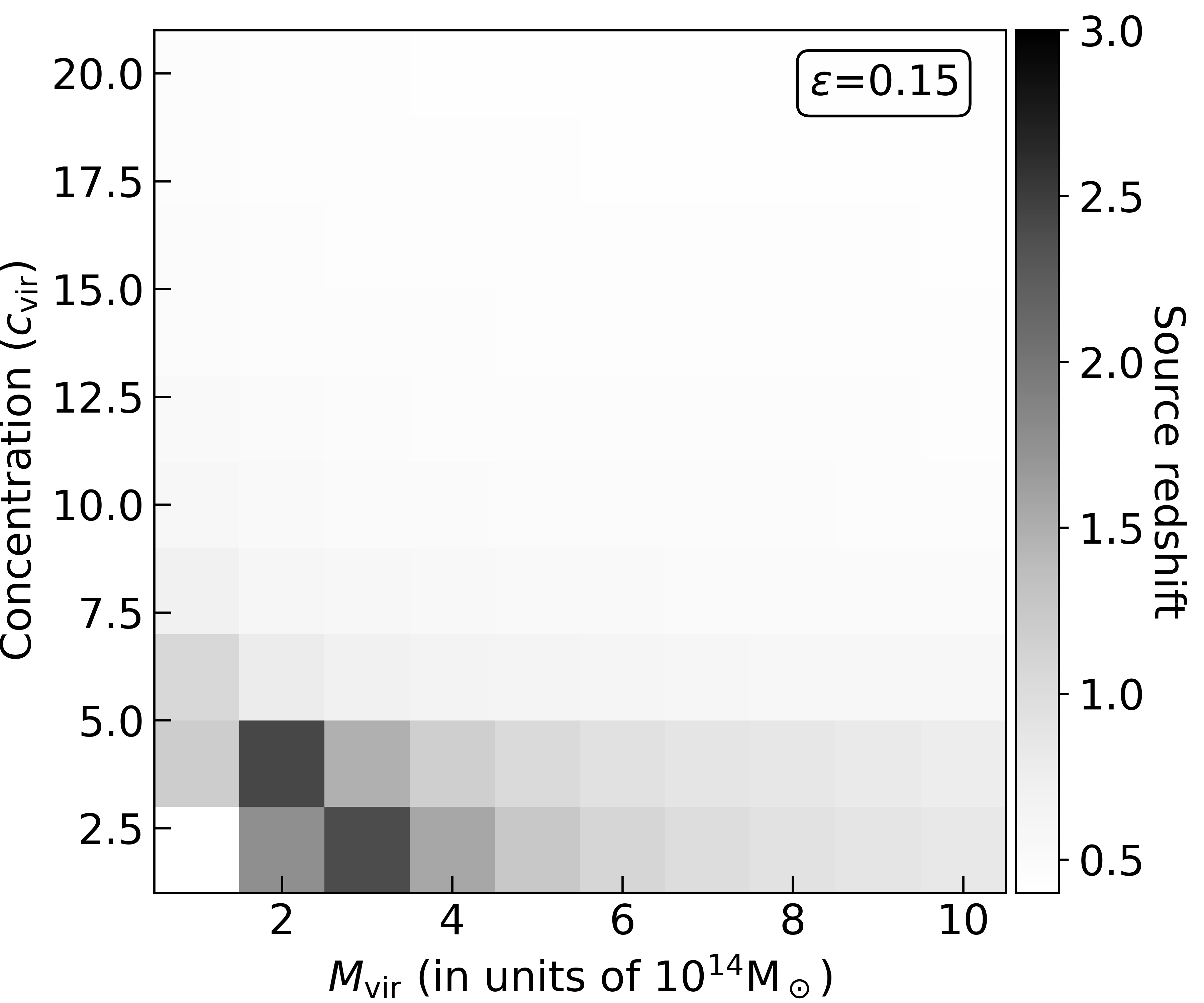}
    \includegraphics[width=8.5cm,height=7.0cm]{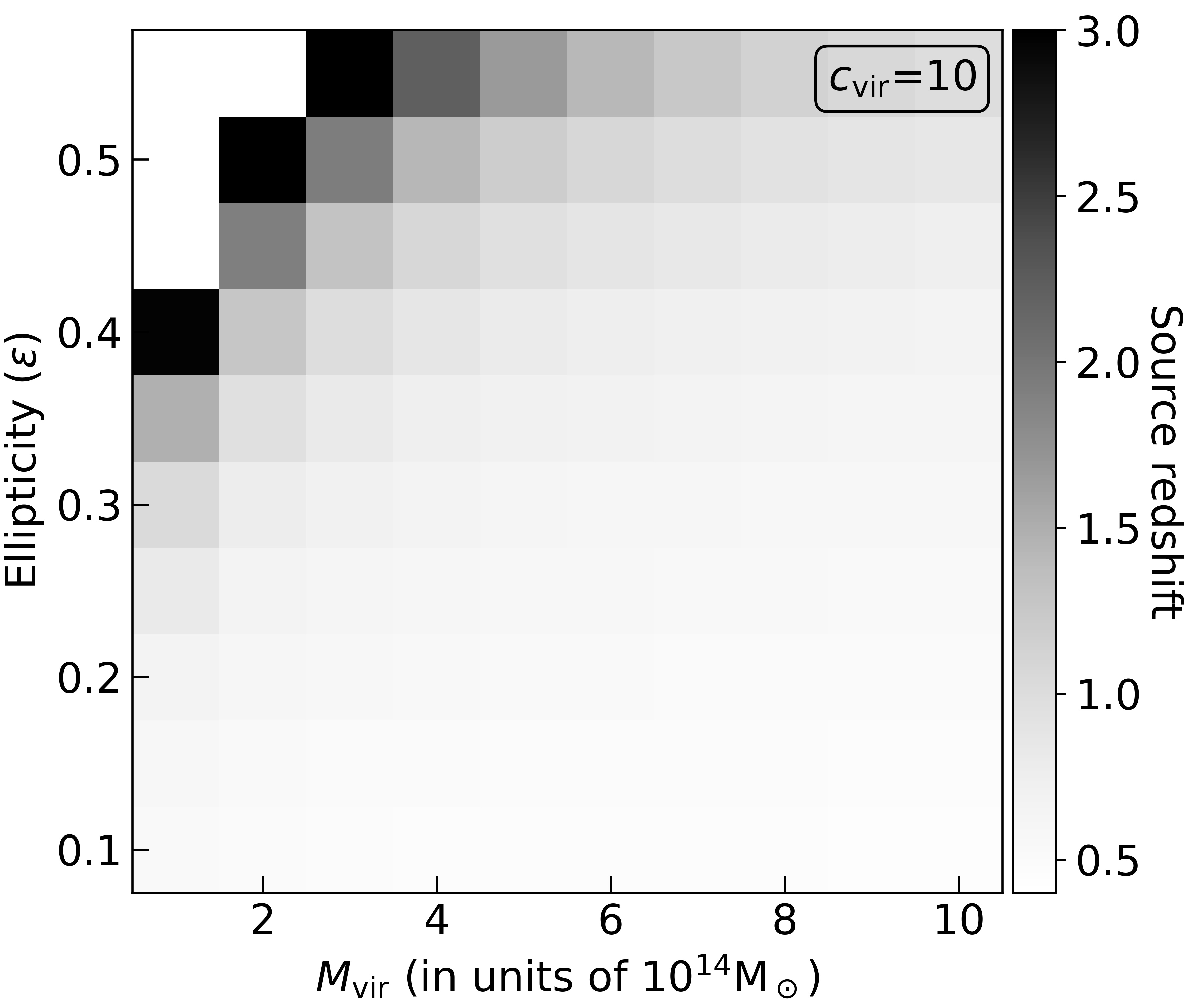}
    \includegraphics[width=8.5cm,height=7.0cm]{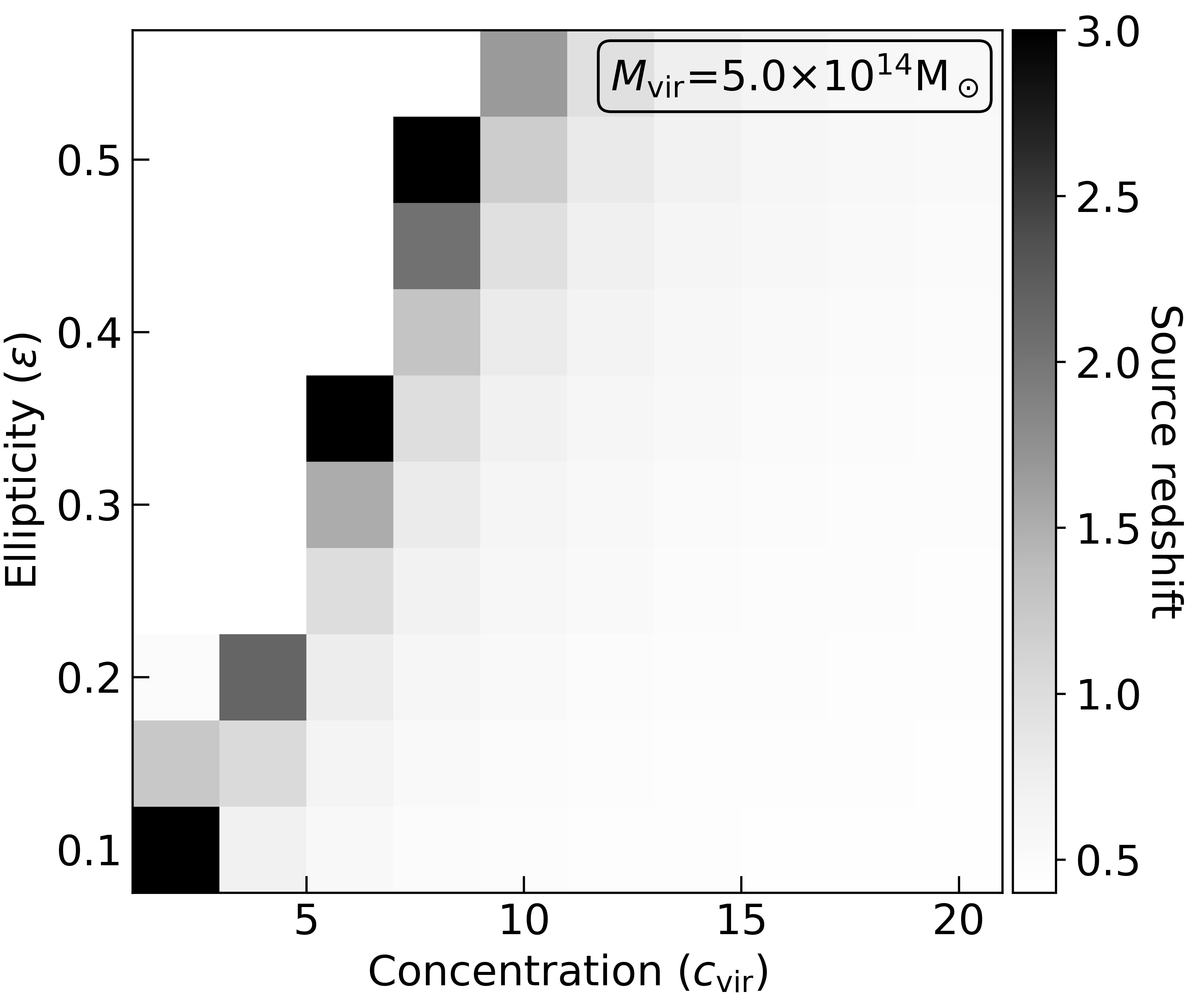}
    \caption{Effect of eNFW lens profile parameters on the HU critical redshift 
    with lens redshift, $z_d=0.4$. From top-to-bottom, we fixed the 
    ellipticity~($\epsilon=0.15$), concentration parameter ($c_{\rm vir}=10$), 
    and total mass ($M_{\rm vir}=5{\times}10^{14}{\rm M_\odot}$) while varying 
    the other two parameters and plot the critical redshift~($z_{\rm hu}$) at 
    which the HU gets critical in each panel, respectively. A sudden change in 
    pixel values near the left edge in each panel represents the lack of HU for 
    those parameters values.}
    \label{fig:VaryHUeNFW}
\end{figure}
%%%%%%%%%%%%%%%%%%%%%%%%%%%%%%%%%%%%%%%%%%%%%%%%%%%%%%%%%%%%%%%%%%%%%%%%%%%%%%%%%%%%%%%%%%%

A source lying near a point singularity leads to a characteristic image formation
with the total signed magnification (of all the images belonging to the characteristic
image formation) summing up to zero~\citep{2009JMP....50c2501A}.
For example, a source lying near a HU leads to a ring-like characteristic image 
formation (off-centered from the lens center) made of four images in the lens plane, 
and the summed magnification of these four images is zero.
The magnification of an image represents the observed flux of the lensed image~($F$) 
divided by the flux of the unlensed source~($F_s$),
\begin{equation}
    \mu = \pm\frac{F}{F_{s}},
\end{equation}
with `+' for minima/maxima images and `-' for saddle-points.
Observations only provide us the lensed image flux and the unlensed source flux 
remains unknown.
Hence, the magnification of an image is not an observable quantity (nor the signed
sum of the magnification, i.e., $\Sigma_i\mu_i$).
However, we can construct an observable quantity known as 
\emph{R-quantity}~\citep{1998MNRAS.295..587M} defined as
\begin{equation}
    R \equiv \frac{\Sigma_i\mu_i}{\Sigma_i|\mu_i|}
    = \frac{\Sigma_i (\pm) F_i}{\Sigma_i F_i} 
    \label{eq:gen_RQ}
\end{equation}
where $\mu_i$ is the magnification and $F_i$ is the observed flux of the $i$-th image. 
For fold, cusp, and HU, the \emph{R-quantity} becomes
\begin{equation}
    \begin{aligned}
        R_{\rm fold} &\equiv \frac{\mu_1+\mu_2}{|\mu_1|+|\mu_2|} = \frac{F_1-F_2}{F_1+F_2}, \\
        R_{\rm cusp} &\equiv \frac{\mu_1+\mu_2+\mu_3}{|\mu_1|+|\mu_2|+|\mu_3|} = 
        \frac{F_1-F_2+F_3}{F_1+F_2+F_3}, \\
        R_{\rm hu}   &\equiv \frac{\mu_1+\mu_2+\mu_3+\mu_4}{|\mu_1|+|\mu_2|+|\mu_3|+|\mu_4|} 
        = \frac{F_1-F_2+F_3-F_4}{F_1+F_2+F_3+F_4}.
    \end{aligned}
    \label{eq:spec_RQ}
\end{equation}
We note that in our definition, the \emph{R-quantity} can be both positive and 
negative.
In principle, just to see the deviation of \emph{R-quantity} from zero, one does 
not have to worry about the sign. 
However, as our main goal is to understand the effect of substructures on 
the~$R_{\rm hu}$, it would be interesting to observe the effect of substructure(s) 
when they lie near positive and negative parity images.
In addition, earlier studies showed that substructures tend to suppress saddle 
points more than amplifying the 
minima~\citep[e.g.,][]{2002ApJ...580..685S, 2004A&A...423..797B, 2004ApJ...610...69K}, 
implying that the skewness in the distribution of \emph{R-quantity} around zero 
may allow us to gain further knowledge about the lens system properties.

In our current work, we focus on understanding the effect of eNFW lens parameters 
and substructures only on~$R_{\rm hu}$, and we leave similar studies for other
singularities for future work. 
For simplicity, we only consider point sources throughout this work. 
For a given substructure, replacing the point source with an extended one will 
suppress the effect of the substructure depending on the size of the source.

%%%%%%%%%%%%%%%%%%%%%%%%%%%%%%%%%%%%%%%%%%%%%%%%%%%%%%%%%%%%%%%%%%%%%%%%%%%%%%%%%%%%%%%%%%%
\begin{figure*}
    \centering
    \includegraphics[scale=0.58]{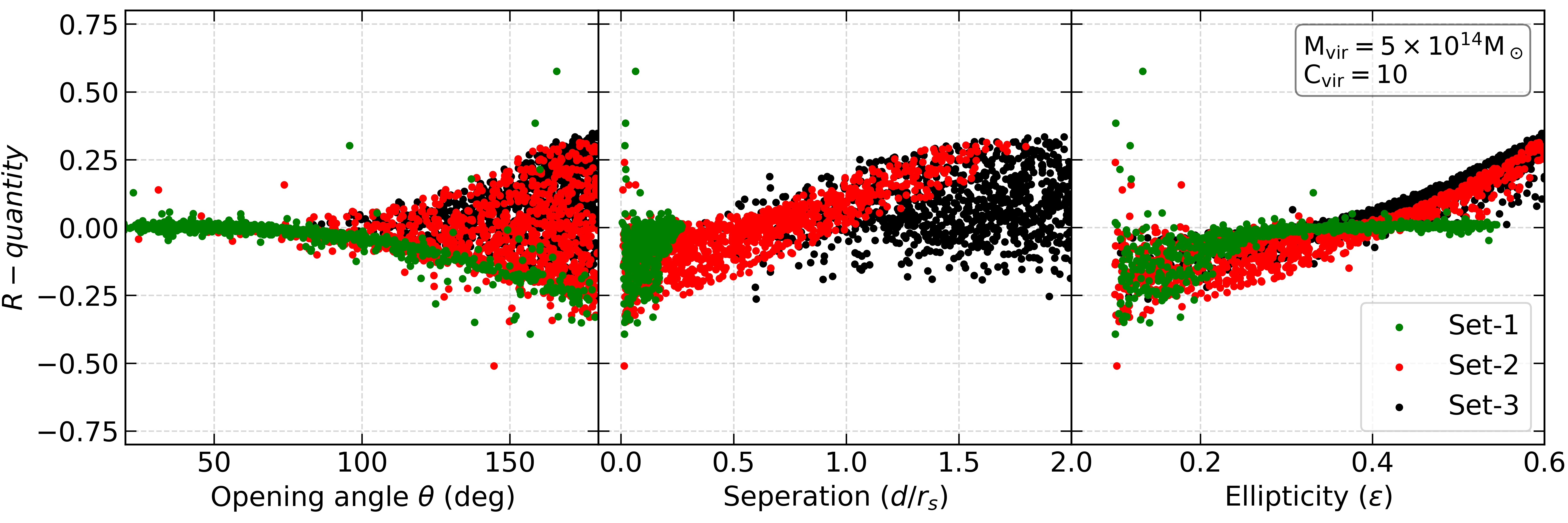}
    \caption{\Rq as a functions of ellipticity~($\epsilon$) for an eNFW lens. In 
    each panel, green, red, and black points represent the lens systems belonging 
    to set-1~(where \Rq is essential represents~$R_{\rm hu}$), set-2, and set-3, 
    respectively. In the left, middle, and right panels, the \emph{R-quantity} is 
    plotted as a function of image opening angle~($\theta$), maximum image 
    separation~($d$; in units of~$r_s$), and lens ellipticity~($\epsilon$), respectively.} 
    \label{fig:oneNFW_eps_dist}
\end{figure*}
%%%%%%%%%%%%%%%%%%%%%%%%%%%%%%%%%%%%%%%%%%%%%%%%%%%%%%%%%%%%%%%%%%%%%%%%%%%%%%%%%%%%%%%%%%%

%%%%%%%%%%%%%%%%%%%%%%%%%%%%%%%%%%%%%%%%%%%%%%%%%%%%%%%%%%%%%%%%%%%%%%%%%%%%%%%%%%%%%%%%%%%
\begin{figure}
    \centering
    \includegraphics[scale=0.55]{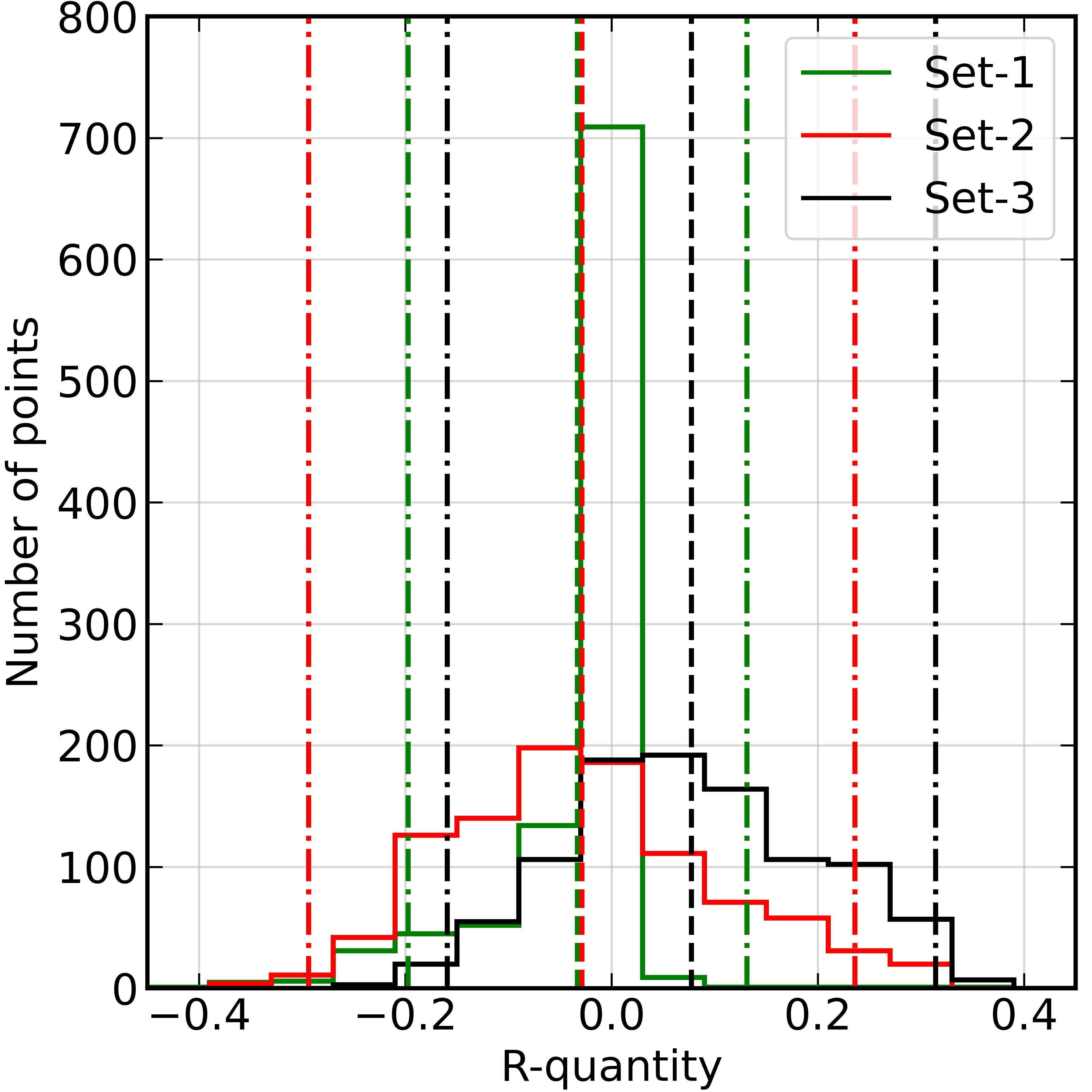}
    \caption{\emph{R-quantity} histogram plot corresponding to Figure~\ref{fig:oneNFW_eps_dist}.
    The green, red, and black histograms correspond to set-1, set-2, and set-3, shown
    by the same color points in Figure~\ref{fig:oneNFW_eps_dist}. The vertical dashed
    lines are the mean value, and the dashed-dotted lines cover the~$2\sigma$ region
    around the mean. i.e.,~[mean-$2\sigma$, mean+$2\sigma$].}
    \label{fig:oneNFW_eps_hist}
\end{figure}
%%%%%%%%%%%%%%%%%%%%%%%%%%%%%%%%%%%%%%%%%%%%%%%%%%%%%%%%%%%%%%%%%%%%%%%%%%%%%%%%%%%%%%%%%%%

%%%%%%%%%%%%%%%%%%%%%%%%%%%%%%%%%%%%%%%%%%%%%%%%%%%%%%%%%%%%%%%%%%%%%%%%%%%%%%%%%%%%%%%%%%%
\section{Image formation near HU}
\label{sec:ImageHU}

By definition, for a given lens, we can only observe the image formation near HU if 
the caustic network shows an exchange of cusp between radial and tangential caustic.
Hence, the simplest lens model that can give rise to a HU singularity is a 
\textit{non-singular elliptical} lens~\citep{1992grle.book.....S, 2020MNRAS.492.3294M}.
In an ideal scenario, a source lying near a HU gives rise to a ring-like image 
formation made of four images and one extra image.
However, even for the simplest lens models, we can see a large variation in the 
observed image formation near HU.

Examples of image formation near the HU singularity for an elliptical Navarro–Frenk–White~(eNFW) 
lens (Appenddix~\ref{app:LensNFW}) are shown in Figure~\ref{fig:ImageHU}.
The virial mass~($M_{\rm vir}$), concentration parameter~($c_{\rm vir}$) and 
ellipticity~($\epsilon$) of the lens are fixed to $10^{14}~{\rm M_\odot}$, 10, and 0.3, respectively.
In the top row, the source is placed at redshift, $z_s=1.0~(=z_{\rm hu})$.
The left and right panels represent the source and lens planes, respectively.
In the left (right) panel, the solid and dashed curves represent the tangential and 
radial caustics (critical curves), respectively.
In the left panel, we can see (more clearly in the inset plot) that 
along the y-axis, the radial and tangential 
caustics are meeting with each other at the cusp points, which is a sign of HU 
singularity getting critical, whereas in the right panel, the radial and tangential critical 
curves meet with each other along the y-axis.
In the left and right panels, the black dot, green diamond, and red star
represent the source positions in the source plane and the corresponding image 
positions in the lens plane, respectively.
For the green source, we can see the ring-like formation made of four images on 
the negative y-axis, and the fifth (global minima) image lies on the positive y-axis.
This image formation can be considered as an `ideal' HU image formation.
When we move the source relatively far from the cusp point (black dot in the top 
row) on the y-axis, we can see that the corresponding ring-like image formation 
gets deformed, and three images inside the critical curves form an arc-like 
structure with the arc being opened away from the centre of the lens.
In addition, if we move the source away from the y-axis (represented by the red star), 
we observe another possible deformation in the characteristic HU image formation
where two of the images lie very close to each other again forming an arc-like  
structure\footnote{One can also study such a pair separately using \emph{R-quantity} 
corresponding to fold, i.e., $R_{\rm fold}$. We refer readers 
to~\citet{2009JMP....50c2501A} for an example.} near the centre of the lens 
whereas the other two images remain isolated. 

As mentioned in \citet{2021MNRAS.503.2097M}, the HU ring-like image formation is 
also observable even before and after the HU critical redshift ($z_{\rm hu}$).
In Figure~\ref{fig:ImageHU}, we show image formation for source redshift, $z=0.8~(<z_{\rm hu})$
and $z=2.0~(>z_{\rm hu})$ in the middle and bottom row, respectively.
We can see that the image formations in both the middle and bottom rows for 
both black and green sources are similar to the image formations in the top panel.
The red star in the bottom panel represents the image formation similar to the
one detected for system-2 in~\citet{2023arXiv230309568L}.
However, the area covered by the four images in the source plane is 
decreased~(increased) if $z<z_{\rm hu}$ ($z>z_{\rm hu}$). 
We can understand this from the fact that for a given strong lens, the size of 
caustics and critical curves increase with increasing source redshift.
Hence, for a source redshift $z<z_{\rm hu}$ ($z>z_{\rm hu}$), the area covered
by the critical curves will be smaller (larger) than~$z=z_{\rm hu}$ case.
The above image formations near the HU show a very small set of possible geometries 
of what we can observe in actual cluster lensing observations as we are only 
considering an ideal lens without any perturbations and a point source.
One possible way to get a complete catalog of realistic image formations near HU 
is to manually lens deep fields by actual cluster lenses as done in~\citet{2022MNRAS.515.4151M}.

At this stage, one can ask the question,
\emph{when the image formation should be treated as characteristic HU-like image 
formation?} 
Of course, for an image formation to be considered as the characteristic HU (-like) 
image formation structure, the source redshift should be within a specific range 
around $z_{\rm hu}$, i.e., $[z_{\rm hu}-\Delta z, z_{\rm hu}+\Delta z]$ but
(according to our knowledge) we do not have an analytical relation that can 
help us determine the redshift range. 
The other problem is that even in a given redshift range around $z_{\rm hu}$, we can 
get various image formations by continuously varying the source position. 
Again, (according to our knowledge) we do not have any analytical relation which tells 
us where the source must lie so that the corresponding image formation can be considered 
as characteristic HU-like image formation. 
Although, we can put a lower limit on the total magnification of the source to 
constrain the source position around the caustics.
Instead, if we ask, \emph{Is the alleged HU-like image formation useful to extract 
any information about the lens or source?}, we might be able to narrow down the
range of various lens system parameter values that are useful when we observe an
HU-like image formation.
To investigate the usability of HU-like image formation to detect substructure near
any of the image, we study the dependence of~$R_{\rm hu}$ on various lens system
parameters.
To quantify the HU-image formation, following~\citet{2003ApJ...598..138K}, we also
introduce two additional parameters, $\theta$, the opening angle of HU-like image 
formation with respect to the lens center and $d$, the maximum separation between 
the images of HU-like image formation.
In the following sections, we vary the lens properties and source position in 
the source plane and study how the $R_{\rm hu}$ varies as a function of $(\theta,d)$.
Similar to \citet{2021MNRAS.503.2097M}, we fix the redshift range $\Delta z$ 
around the $z_{\rm hu}$ such that  $\Delta z = 0.1 \times a(z_{\rm hu})$, where 
$a(z_{\rm hu})$ is the distance ratio at $z_{\rm hu}$.
Outside this redshift range the image formation will gradually change
into the generic five image formation.
For an eNFW lens, the above choice of~$\Delta z$ value is well supported from 
Figure~\ref{fig:ImageHU} where~$z_{\rm hu}=1.0$ and $\Delta z = 0.1 \times a(z_{\rm hu})$
implies a redshift range~$\sim[0.87,1.2]$. 
However, we can observe HU-like image formation in a larger redshift range~[0.8, 2.0].

%%%%%%%%%%%%%%%%%%%%%%%%%%%%%%%%%%%%%%%%%%%%%%%%%%%%%%%%%%%%%%%%%%%%%%%%%%%%%%%%%%%%%%%%%%%
\begin{figure*}
    \centering
    \includegraphics[scale=0.57]{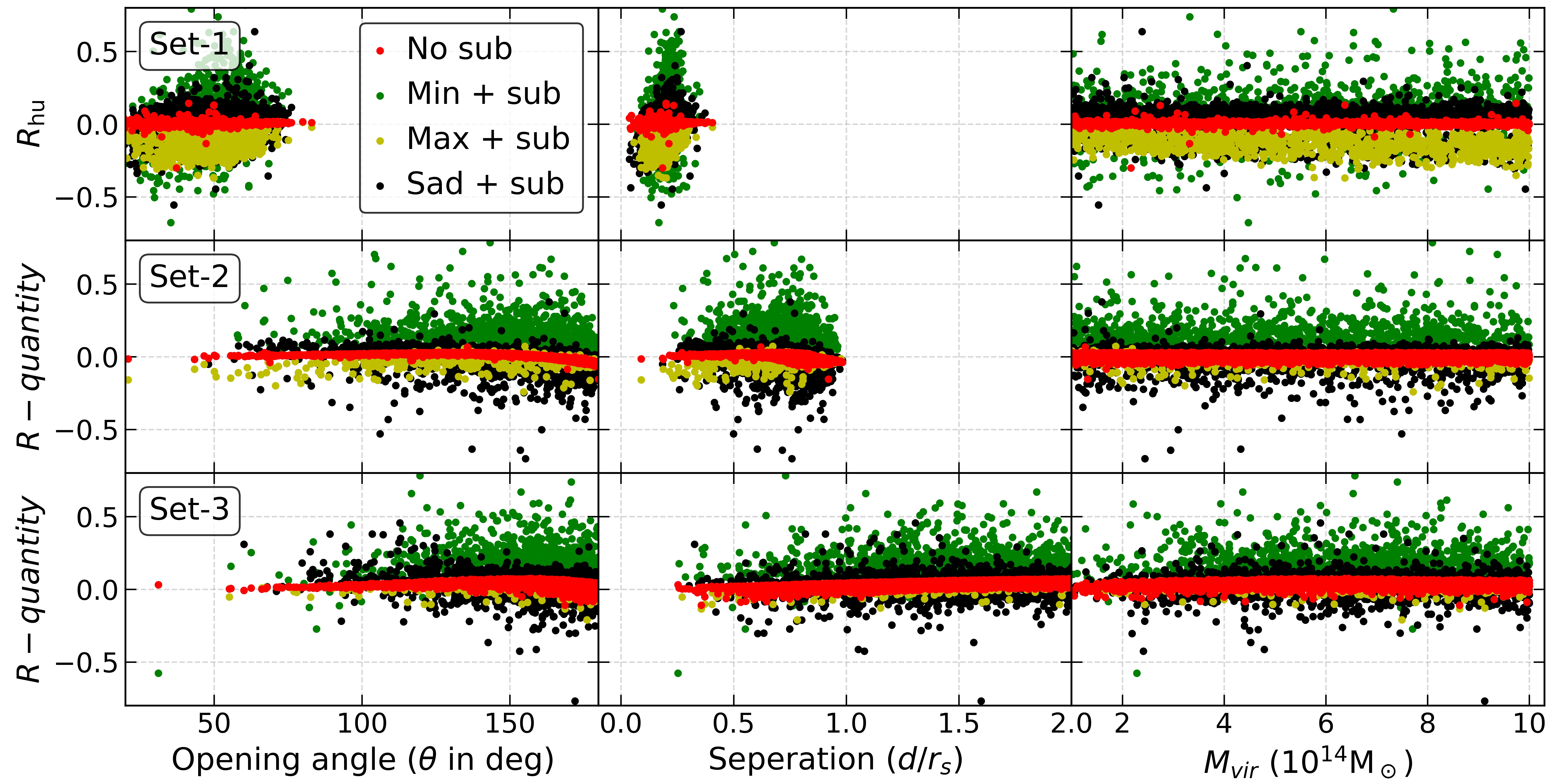}
    \caption{Effect of a substructure on~$R_{\rm hu}$~(\Rq) for eNFW lens. In the top row, red 
    points represent the~$R_{\rm hu}$ in the absence of substructure while varying the lens 
    mass~($M_{\rm vir}$). The left, middle, and right panels plot the~$R_{\rm hu}$ as 
    a function of opening angle~($\theta$), image separation~($d$), and lens 
    mass~($M_{\rm vir}$). The green, yellow, and black points show~$R_{\rm hu}$ in the 
    presence of a substructure (in the mass range~[$10^7~{\rm M_\odot}$,~$10^{10}~{\rm M_\odot}$]) 
    near minima, maxima and saddle-points in the characteristic image formation, respectively. 
    Similarly, in the middle and bottom rows, we plot the~\Rq for set-2 and set-3 in the 
    absence and presence of the substructure.}
    \label{fig:eNFWSub_mvir_dist}
\end{figure*}
%%%%%%%%%%%%%%%%%%%%%%%%%%%%%%%%%%%%%%%%%%%%%%%%%%%%%%%%%%%%%%%%%%%%%%%%%%%%%%%%%%%%%%%%%%%

%%%%%%%%%%%%%%%%%%%%%%%%%%%%%%%%%%%%%%%%%%%%%%%%%%%%%%%%%%%%%%%%%%%%%%%%%%%%%%%%%%%%%%%%%%%
\begin{figure*}
    \centering
    \includegraphics[scale=0.57]{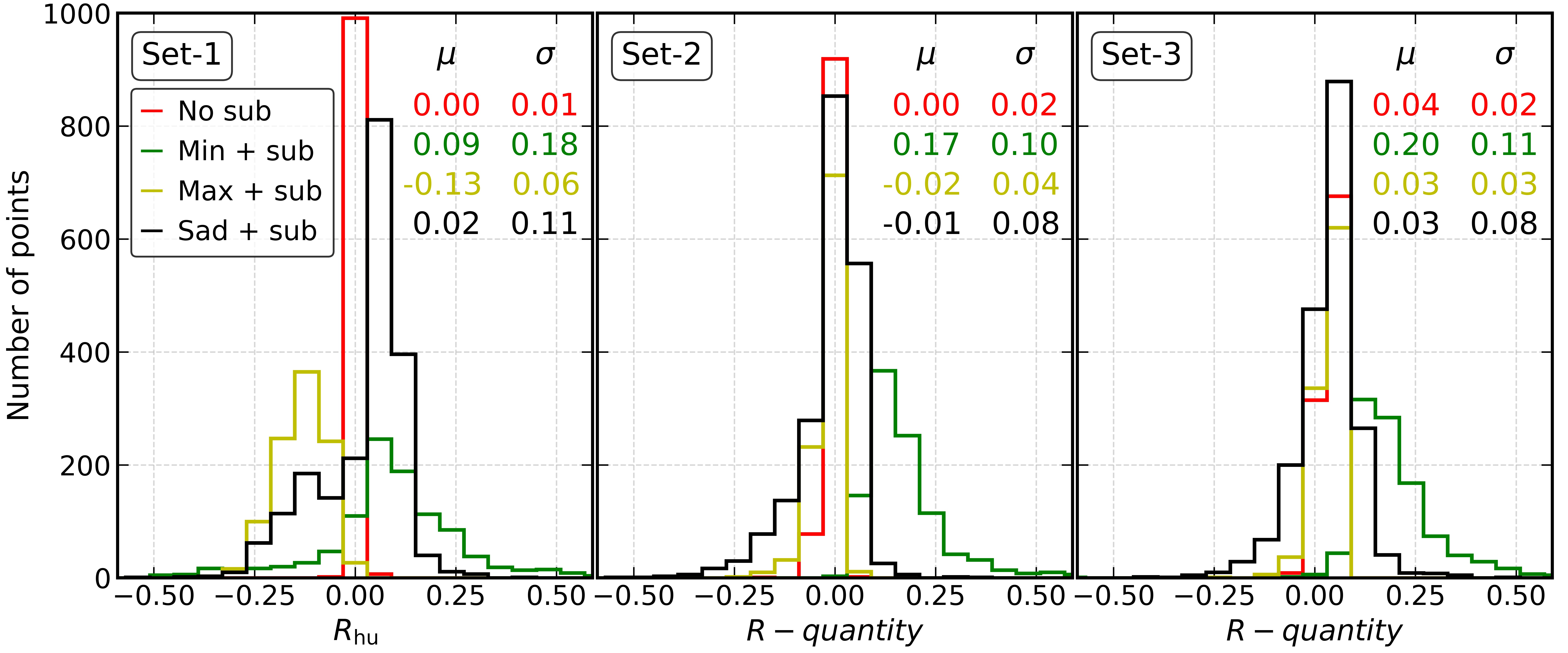}
    \caption{Histogram plot for simulated systems shown in Figure~\ref{fig:eNFWSub_mvir_dist}. 
    Left, middle, and right panels are corresponding to set-1, set-3, and set-3, 
    respectively. Similar to Figure~\ref{fig:eNFWSub_mvir_dist}, the red, green,
    yellow, and black histograms correspond to no substructure, substructure
    near minima, substructure near maxima, and substructure near one of the saddle
    points, respectively. For each histogram, the mean~($\mu$) and standard 
    deviation~($\sigma$) are shown in the upper right part of each panel. We stress
    that in these plots~$\mu$ does not represent magnification but represents the 
    average value.}
    \label{fig:eNFWSub_mvir_hist}
\end{figure*}
%%%%%%%%%%%%%%%%%%%%%%%%%%%%%%%%%%%%%%%%%%%%%%%%%%%%%%%%%%%%%%%%%%%%%%%%%%%%%%%%%%%%%%%%%%%

%%%%%%%%%%%%%%%%%%%%%%%%%%%%%%%%%%%%%%%%%%%%%%%%%%%%%%%%%%%%%%%%%%%%%%%%%%%%%%%%%%%%%%%%%%%
\section{\lowercase{e}NFW Lens}
\label{sec:OneNFW}

In our current work, we only focus on the eNFW lens profile~(Appendix~\ref{app:LensNFW}).
The use of eNFW profile is motivated considering the fact that it is widely used to 
model dark matter halos from galaxy lenses to galaxy-cluster 
lenses~\citep[e.g.,][]{2003MNRAS.344.1029D, 2009MNRAS.392..945V, 2010MNRAS.405.2215O,
2012ApJ...750...10S, 2014A&A...572A..19F}.
An eNFW profile can be described using three parameters: ellipticity ($\epsilon$),
virial mass~($M_{\rm vir}$), and concentration parameter~($c_{\rm vir}$).
Varying any of these parameters will change the source redshift at which the lens will
start to behave as a strong lens, i.e., the formation of critical curves and caustics, 
which in turn also affect the corresponding critical HU redshift~($z_{\rm hu}$) making it 
important to understand how~$z_{\rm hu}$ depends on various eNFW parameters.
Hence, in Section~\ref{ssec:critz_enfw}, we investigate the dependency of~($z_{\rm hu}$)
on various eNFW parameters.
After that, in Section~\ref{ssec:rhu_enfw_eps} and~\ref{ssec:rhu_enfw_mcvir}, we 
study the dependence of HU magnification relation on~$\epsilon$ 
and~($M_{\rm vir}$, $c_{\rm vir}$) for an isolated eNFW lens, respectively.
Finally, in Section~\ref{ssec:rhu_enfw_ee}, we briefly discuss the effect of external shear
on HU magnification relation.

%%%%%%%%%%%%%%%%%%%%%%%%%%%%%%%%%%%%%%%%%%%%%%%%%%%%%%%%%%%%%%%%%%%%%%%%%%%%%%%%%%%%%%%%%%%
\subsection{$z_{\rm hu}(\epsilon, M_{\rm vir}, c_{\rm vir})$}
\label{ssec:critz_enfw}

In Figure~\ref{fig:VaryHUeNFW}, we plot the critical HU redshift~($z_{\rm hu}$) as
a function of~$(\epsilon, M_{\rm vir}, c_{\rm vir})$\footnote{We refer
reader to~\citet{2020MNRAS.492.3294M} on discussion about how to calculate~$z_{\rm hu}$.}.
In the top, middle, and bottom panels, we fix $\epsilon=0.15$, $c_{\rm vir}=10$, 
$M_{\rm vir}=5\times10^{14}~{\rm M_\odot}$, respectively and vary the other two parameters.
In the top panel, we observe that for a given ellipticity value ($\epsilon=0.15$), 
$c_{\rm vir}$ values above 10 lead to $z_{\rm hu}$ very close to the lens redshift~($z_l$)
for all lens mass values.
This can be understood from the fact that for a given lens mass an increase in $c_{\rm vir}$ 
decreases the corresponding scale radius~$r_s$ which makes the central region critical
for strong lensing for smaller source redshift.
As a result, the lens shows the cusp exchange between tangential and radial caustics at
smaller source redshifts.
The opposite happens for smaller values of the $c_{\rm vir}$, and we observe the HU getting 
critical at redshifts around one.
On the other hand, if we fix the~$c_{\rm vir}=10$ and vary ($M_{\rm vir}$, $\epsilon$) 
as shown in the middle panel of Figure~\ref{fig:VaryHUeNFW}, we notice that $z_{\rm hu}$ 
increases with an increase in~$\epsilon$ for all mass values.
This implies that re-plotting the top panel for higher~$\epsilon$ values will lead to an
vertical shift.
From the bottom panel in Figure~\ref{fig:VaryHUeNFW}, we see that for a given lens mass, 
the effect of increasing~$c_{\rm vir}$ can be compensated by increasing the ellipticity 
as the higher value of~$c_{\rm vir}$ will decrease the $z_{\rm hu}$ whereas an increase 
in~$\epsilon$ will increase the~$z_{\rm hu}$.

If the~$z_{\rm hu}$ lies close to the $z_l$ (implying smaller distance ratio, $a$) then 
the cross-section to observe HU image formation decreases significantly because of two reasons: 
(i) for smaller distance ratio, the caustic curves evolve rapidly as we increase the
source redshift implying a smaller redshift range for HU-like image formation,
(ii) for a smaller distance ratio, the area covered by the caustics in the source plane
is also small, implying fewer chances of encountering a source for HU-like image formation.
Figure~\ref{fig:VaryHUeNFW} allows us to get an idea about the preferred range of eNFW
lens parameters for HU-like image formation.
We observe that the more the elliptical lens is, the more chances we have of observing
an HU-like image formation.
On the other hand, if lens has large~$c_{\rm vir}$ then probability of HU-like image
formation decreases as the $z_{\rm hu}$ will be close to the~$z_l$.

%%%%%%%%%%%%%%%%%%%%%%%%%%%%%%%%%%%%%%%%%%%%%%%%%%%%%%%%%%%%%%%%%%%%%%%%%%%%%%%%%%%%%%%%%%%
\subsection{\emph{R-quantity} vs. ellipticity~($\epsilon$)}
\label{ssec:rhu_enfw_eps}

To understand the effect of eNFW lens ellipticity~($\epsilon$) on the HU magnification
relation, we simulate three sets of~$10^3$ lens systems with lens mass~($M_{\rm vir}$) 
equal to~$5\times10^{14}{\rm M}_\odot$, concentration parameters ($c_{\rm vir}$) 
equal to~10 and vary ellipticity~($\epsilon$) in the range~[0.1, 0.6]. 
In set-1, we fix the source redshift in the range mentioned above, 
i.e.,~$z_{\rm hu} \pm \Delta z$, and we put the source in a circle of~5~Kpc around 
the cusp point in the source plane.
In set-2, we fix the source redshift in the same range, but the source can 
lie anywhere in the five-image region in the source plane.
In set-3, the source can lie anywhere in the five-image region at any redshift 
equivalent to a set of generic five-image formations.
In each case, we remove the global minima image to calculate 
the HU magnification relation
from the remaining four images, as these images are (in principle) part of the 
characteristic HU image formation.
For set-1, magnification relation is denoted by~$R_{\rm hu}$ whereas for set-2 and 
set-3 the magnification relation is denoted by~\Rq throughout this manuscript.

The corresponding results are shown in Figure~\ref{fig:oneNFW_eps_dist}.
In the left, middle, and right panels, we plot~\Rq as a function of the 
opening angle~($\theta$), maximum image separation~($d$; in units of scale 
radius~$r_s$), and ellipticity~($\epsilon$).
In the left panel, for set-1 (green points), we observe that the average 
value of \Rq (which is equivalent to~$R_{\rm hu}$ for set-1) moves towards 
negative values as we increase~$\theta$ without any significant scatter in 
the values.
In the middle panel, all the points lie below~$d=0.2$, which is expected as 
all images form sufficiently close to each other due to the source lying near
the caustics in the source plane.
In the right panel, we notice that smaller values of~$\epsilon$ leads to 
more scatter in~$R_{\rm hu}$ with an average value less than zero, and as we 
increase~$\epsilon$, the scatter decreases with an average value very close 
to zero. 
The behaviour of green points in all panels can be understood by focusing on
the right panel.
For smaller~$\epsilon$, the~$z_{\rm hu}$ is relatively close to the~$z_l$ 
which can have a large~$\theta$ value but smaller~$d$ values as all images 
lie close to the lens centre.
In such cases, the large value of~$\theta$ implies that the maxima image might 
lie very close to the centre (getting further de-magnified), while the two 
saddle images carry roughly similar magnifications leading to a 
negative~$R_{\rm hu}$ value.
The larger scatter around negative~$R_{\rm hu}$ is most likely a result of 
our finite grid resolution to estimate the image position.

%%%%%%%%%%%%%%%%%%%%%%%%%%%%%%%%%%%%%%%%%%%%%%%%%%%%%%%%%%%%%%%%%%%%%%%%%%%%%%%%%%%%%%%%%%%
\begin{figure*}
    \centering
    \includegraphics[scale=0.53]{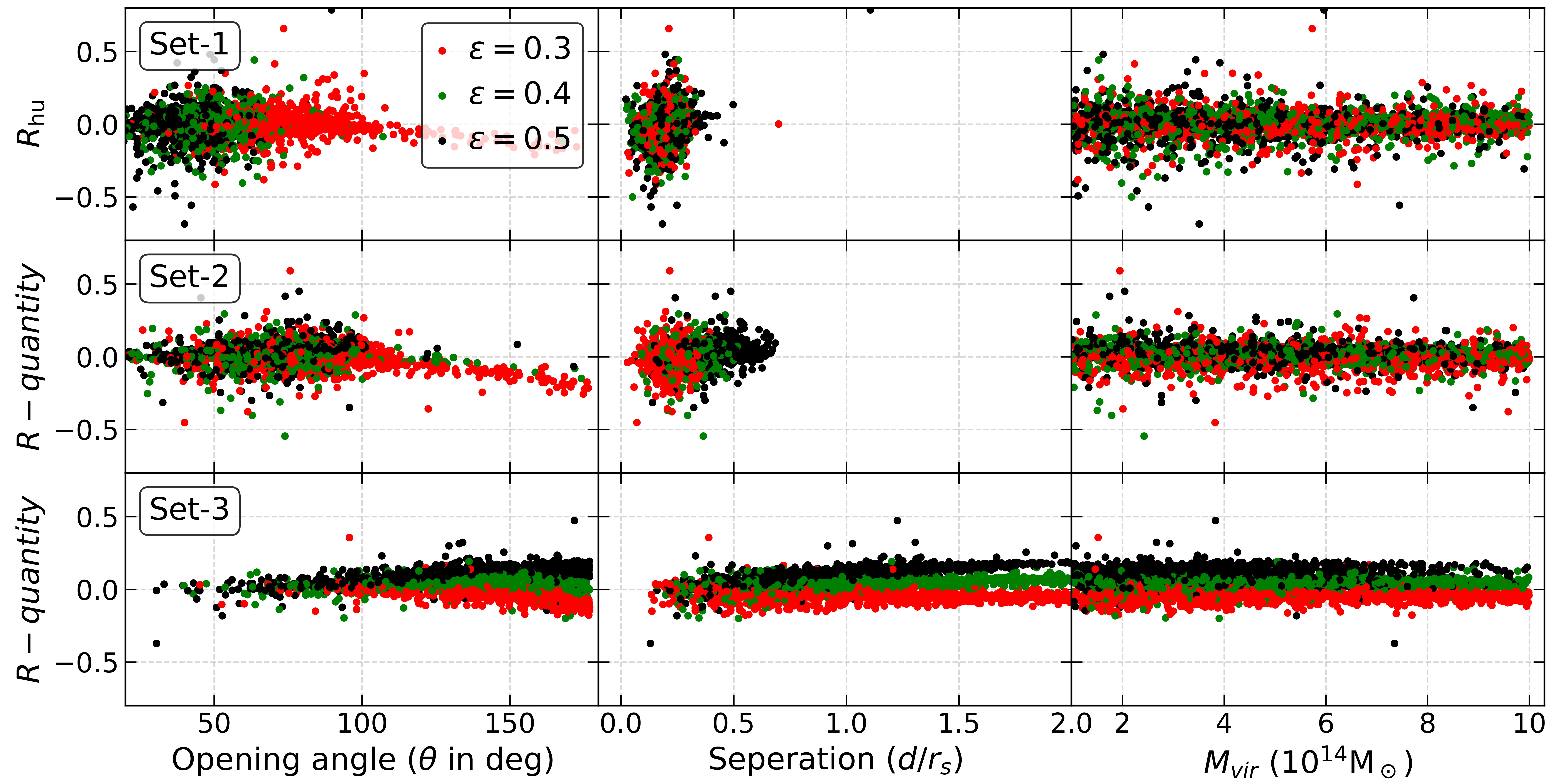}
    \caption{\Rq distribution as a function of the lens mass 
    with~$(\epsilon, c_{\rm vir})=(0.4, 10)$ for eNFW + Many substructures lens. In the top row, 
    red, green, and black points represent the~\Rq values for set-1~(equal to $R_{\rm hu}$)  
    for~$\epsilon=0.3, 0.4, {\rm and}~0.5$ while varying the lens mass~($M_{\rm vir}$). The left,
    middle, and right panels show the~$R_{\rm hu}$ as a function of opening angle~($\theta$),
    image separation~($d$), and lens mass~($M_{\rm vir}$). In the middle and bottom rows, we 
    show the \Rq values for set-2 and set-3, respectively.}
    \label{fig:nfwMany_mvir_dist}
\end{figure*}

%%%%%%%%%%%%%%%%%%%%%%%%%%%%%%%%%%%%%%%%%%%%%%%%%%%%%%%%%%%%%%%%%%%%%%%%%%%%%%%%%%%%%%%%%%%
\begin{figure*}
    \centering
    \includegraphics[scale=0.53]{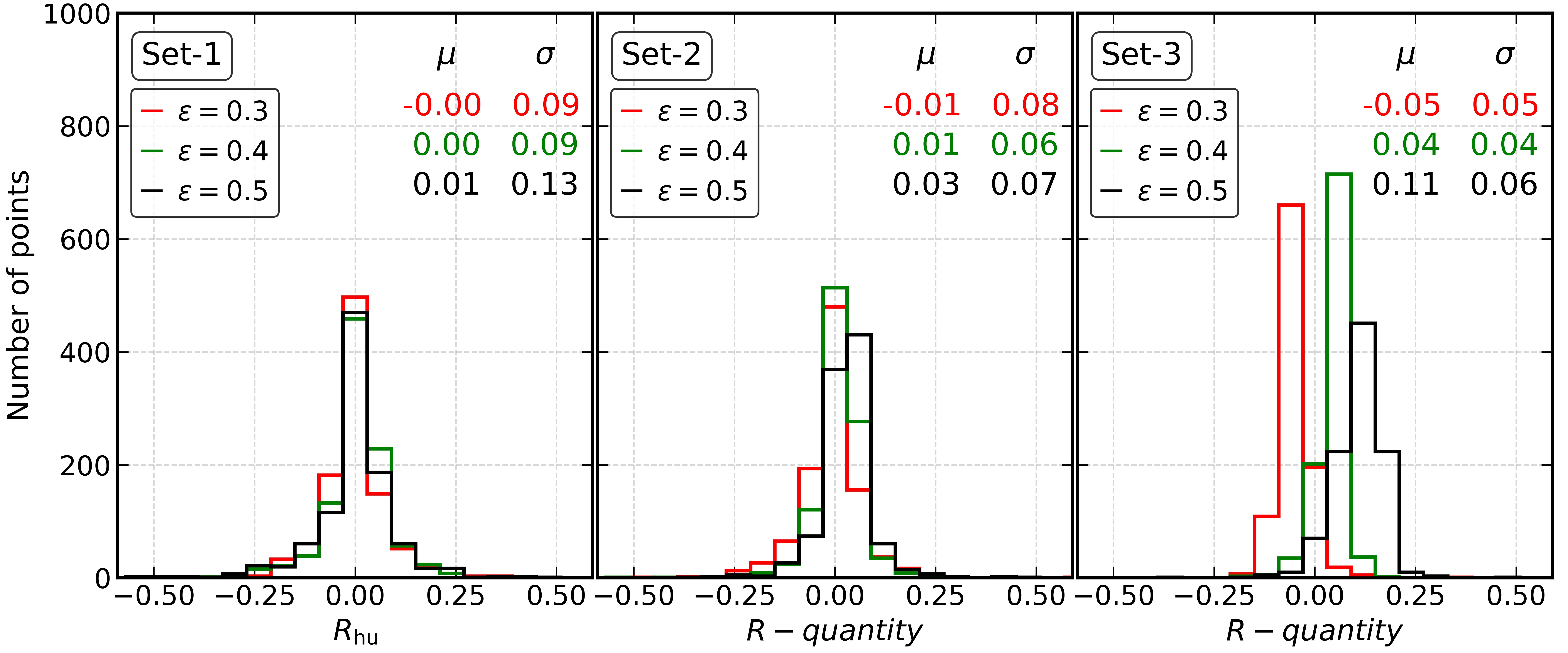}
    \caption{\Rq histogram plot for systems shown in Figure~\ref{fig:nfwMany_mvir_dist}.
    Left, middle, and right panels are corresponding to set-1, set-2, and set-3, 
    respectively. Similar to Figure~\ref{fig:nfwMany_mvir_dist}, the red, green,
    and black histograms correspond to~$\epsilon=0.3, 0.4,{\rm and}~0.5$, 
    respectively. For each histogram, the mean~($\mu$) and standard 
    deviation~($\sigma$) are shown in the upper right part of each panel.}
    \label{fig:nfwMany_mvir_hist}
\end{figure*}

In set-2 (represented by red points), as we remove the constraint on the source 
position, the source can lie in a larger region.
This leads to an increase in the maximum separation~($d$) in the middle panel.
In addition, we also observe a trend in the \Rq values, i.e., an increase 
in~$d$ moves \Rq values from negative to positive values.
The same trend in \Rq can also be seen in the right panel.
Again the large scatter in red points for smaller~$\epsilon$ or~$d$ can 
be explained from the fact that for small~$\epsilon$, the HU gets critical for 
smaller source redshift, implying the corresponding image formation will take 
place near the lens centre, leading to the large scatter in \Rq with smaller~$d$.
The other interesting point to notice is that for~$\epsilon < 0.4$, \Rq prefers 
negative values with a transition to positive values at~$\epsilon \sim 0.4$.
From observations, the mean ellipticity value for cluster scale lenses 
is~$\sim0.46$~\citep{2010MNRAS.405.2215O, 2020MNRAS.496.2591O}, implying that for 
a source in the HU redshift range will lead to image formation with \Rq values 
very close to zero even when the source is sufficiently far from the caustics.
Similarly, for early-type galaxy scale lenses, the observed mean ellipticity value 
is~$\sim0.3$~\citep[e.g.,][]{2008MNRAS.388.1321P} implying, in general, negative
\Rq values with a larger scatter.
Above~$\epsilon > 0.4$, the~\Rq prefers positive values which can be understood
from the fact that for large ellipticities, the saddle-points will lie closer
to the lens centre and get de-magnified similar to the central image leading to
an overall positive~\Rq.
We stress that the above inferences are only valid for isolated galaxy/cluster 
scale lenses (i.e., no substructure) modelled using an eNFW lens profile.
The effect of substructure(s) on~$R_{\rm hu}$ value is explored in the following 
sections.

In set-3 (represented by black points), we relaxed the constraints on both the 
source redshift and source position.
For set-3, we see similar trends as set-2  with~$\theta$ and~$\epsilon$ but now 
we have more systems with large~$d$ values.
Observing the behavior of black points in the right panel, we can infer that for
a given~$\epsilon$, the \Rq value does not vary much, but the corresponding~$d$
value has a larger allowed range compared to set-2.
This implies that for~$\epsilon\gtrsim0.3$, even the generic image 
formation~(set-3) leads to similar \Rq values to when source redshift lies very 
close to the~$z_{\rm hu}$ anywhere in the five-image region in the source 
plane~(set-2).
However, it is not expected to be useful to do flux-ratio analysis
to look for substructures.
Assuming cluster scale lens, in generic five-image formations, the galaxy scale
lenses can affect each image differently which will introduce a scatter in~\Rq
which is not shown in the plot. 
In addition to estimate flux ratios, we need to identify knots inside the lensed
galaxies but since the magnification is moderate for generic five image formation,
it will be very hard to find knots in such lens systems.
Even if we identify a knot in a generic five image formation, the substructure needs 
to lie very close to the knot~(compared to HU image formation) as the effective 
Einstein radius of a substructure will depend on the background 
magnification~\citep[e.g.,][]{2018ApJ...857...25D} again decrease our chances 
to observe such a system.

%%%%%%%%%%%%%%%%%%%%%%%%%%%%%%%%%%%%%%%%%%%%%%%%%%%%%%%%%%%%%%%%%%%%%%%%%%%%%%%%%%%%%%%%%%%
\subsection{\emph{R-quantity} vs. ($M_{\rm vir}, c_{\rm vir}$)}
\label{ssec:rhu_enfw_mcvir}

In this subsection, we fix ellipticity~($\epsilon$) to three different values 0.3, 
0.4, and 0.5 while varying either lens mass~($M_{\rm vir}$) or concentration 
parameter~($c_{\rm vir}$) and simulate three sets of lens systems similar to 
those above.
Again each set contains~$10^3$ lens systems.
The corresponding plots for \Rq as a function of opening angle, maximum image separation,
and lens mass/concentration are presented in Appendix~\ref{app:LensNFW_MCvirDist}.
We observe that~$R_{\rm hu}$ (Set-1; top rows in Figure~\ref{fig:oneNFW_mvir_dist} 
and~\ref{fig:oneNFW_cvir_dist}) remains very close to zero for any value of lens 
mass or concentration parameter for~$\epsilon=0.3, 0.4, 0.5$ which can also be 
inferred from Figure~\ref{fig:oneNFW_eps_dist} implying that~$R_{\rm hu}$ is 
most sensitive to lens ellipticity.
Similarly, for Set-2 and Set-3 also, the range of \Rq value can be inferred from
Figure~\ref{fig:oneNFW_eps_dist} and variation in lens mass or the concentration
parameter does not lead to a significant change in \Rq value.
Hence, from the above analysis, we can infer that ellipticity~($\epsilon$) is one of 
the crucial parameters that determine the scatter in~\Rq~(or~$R_{\rm hu}$) for an enFW lens.
For typical lenses modeled by eNFW profile with~$\epsilon\in[0.3,0.5]$, we 
expect~$R_{\rm hu}$ to lie very close to zero.

%%%%%%%%%%%%%%%%%%%%%%%%%%%%%%%%%%%%%%%%%%%%%%%%%%%%%%%%%%%%%%%%%%%%%%%%%%%%%%%%%%%%%%%%%%%
\subsection{\Rq vs. external shear}
\label{ssec:rhu_enfw_ee}

In general, strong gravitational lenses also have other halos near them.
The presence of such halos near the main lens introduces additional terms in 
the deflection field. 
The amplitude of these additional terms depends on the distance between the main lens 
and the external halo(s) and has the capability to modify the observed image formation.
From section~5.3 of \citet{2020MNRAS.492.3294M}, we can infer that if external 
effects are strong ($\gamma_{\rm ext}\gtrsim0.3$) they can change the overall singularity 
map of an eNFW lens and affect the critical HU redshift.
This implies that in the presence of strong external effects Figure~\ref{fig:VaryHUeNFW}
will get modified.
On the other hand, \Rq (or~$R_{\rm hu}$) is mainly sensitive to the distance of the 
source from the caustics in the source plane and the distance of the corresponding image
formation from the centre of the lens. 
However, as we see from Figure~\ref{fig:oneNFW_eps_dist}, for~$\epsilon\gtrsim0.3$, 
the~$R_{\rm hu}$ is mainly sensitive to the ellipticity, implying that we do not expect 
external shear to make a significant impact  (unless it is very strong) 
in Figure~\ref{fig:oneNFW_eps_dist} and the inferred results.

%%%%%%%%%%%%%%%%%%%%%%%%%%%%%%%%%%%%%%%%%%%%%%%%%%%%%%%%%%%%%%%%%%%%%%%%%%%%%%%%%%%%%%%%%%%
\section{\lowercase{e}NFW Lens + One (satellite-scale) Substructure}
\label{sec:OneNFWSub}

The above analysis allows us to gain insight into the~$R_{\rm hu}$ values that are allowed 
for an isolated eNFW lens without any substructures near the characteristic image formation.
Introducing a substructure near a lensed image, depending on the lensed image and substructure
properties can make the lensed image brighter, fainter, or can split it into two or more 
images modifying the corresponding~$R_{\rm hu}$ (or, in general, \emph{R-quantity}).
In the presence of a substructure, the lens potential is modified as follows,
\begin{equation}
    \Psi(\theta) = \Psi_{\rm p}(\theta) + \Psi_{\rm s}(\theta),
    \label{eq:PotOneNFWSS}
\end{equation}
where~$\Psi_{\rm p}(\theta)$ represents the main lens potential and $\Psi_{\rm s}(\theta)$
represents the substructure potential.
For simplicity, we model the substructure lens using the \emph{singular isothermal sphere}~(SIS) profile.
The lens potential for the SIS lens is given as
\begin{equation}
    \Psi_{\rm SIS}(\theta) = \theta_{\rm E}|\theta|, \qquad {\rm with} \qquad 
    \theta_{\rm E} = 4\pi\frac{D_{ds}}{D_s}\frac{v_d^2}{c^2},
    \label{eq:sis_ein}
\end{equation}
where~$\theta_{\rm E}$ represents the Einstein angle and~$v_d$ represents the velocity
dispersion of the lens.
The total SIS lens mass within the Einstein radius is given as,
$M(\theta_{\rm E})=\pi v_d^2 D_d \: \theta_{\rm E}/{\rm G}$.

The presence of a substructure leads to the formation of additional critical curves and 
caustics in the lens and source plane, respectively.
The area covered by these additional critical curves in the lens plane depends on the
lens mass and the background magnification at the substructure position due to the main 
lens.
If the substructure lies near a minima image, for an SIS lens, it leads to one
additional critical curve with a diamond-shaped caustic in the source plane.
If the substructure lies near a saddle-point, it leads to two critical curves
in the lens plane and two triangular-shaped caustics in the source plane with a highly
de-magnified region between them.
If the substructure lies sufficiently close to the strongly lensed image such that the
unlensed source position falls inside the substructure caustic, the lensed image will 
be further divided into multiple images.
In such cases, it might be easy to identify the presence of the substructure directly
from the observations.
In addition, as shown in earlier works~\citep[e.g.,][]{2002ApJ...580..685S, 
2004A&A...423..797B, 2004ApJ...610...69K}, the substructures tend to suppress saddle-points 
more than amplifying the minima.
Hence, we can expect that a substructure present near the minima and saddle-point 
will shift the~$R_{\rm hu}$ towards positive values. 

Considering the above points, in our current work, we only simulate lens systems where 
the lensed image is outside the substructure critical curve as well as we divide them 
into different sets based on the image type, i.e., whether the substructure lies near
minima, maxima, or saddle-point.
We randomly choose the substructure mass from the range~$[10^7~{\rm M_\odot},\:10^{10}~{\rm M_\odot}]$. 
This mass range is motivated by the fact that our main 
aim is to test whether the HU image formation can be used to detect ``dark'' substructures  
(i.e., those without significant stellar/baryonic content) and differentiate between 
different dark matter models. A detailed analysis on differentiating different dark matter
models using HU image formation is subjected to our ongoing work and will be presented in
a forthcoming paper.
We draw a circle with a radius five times the Einstein angle of the substructure around 
the lensed images, which are part of the characteristic image formation, and randomly 
place the substructure in it.
Here we note that while drawing the circle around lensed images, we estimate the
Einstein angle using Equation~\eqref{eq:sis_ein} without considering the effect of main
lens.
As we can see in right panel of Figure~\ref{fig:oneNFW_eps_dist}, the~$R_{\rm hu}$ values 
lie very close to zero for~$\epsilon\sim0.4$~(which is also close to observed cluster 
lenses~\citealt{2010MNRAS.405.2215O}), we fix the lens ellipticity to 0.4 in our simulations 
and vary~$c_{\rm vir}$ and~$M_{\rm vir}$.

The~\Rq value for simulated systems with varying~$M_{\rm vir}$ are shown in 
Figure~\ref{fig:eNFWSub_mvir_dist}.
In the top, middle, and bottom rows of Figure~\ref{fig:eNFWSub_mvir_dist}, we show 
the~\Rq for set-1~(essentially equal to~$R_{\rm hu}$), set-2, and set-3, 
respectively.
The red, green, yellow, and black points in all panels represent the~\Rq values for cases 
when there is no substructure, a substructure near minima, a substructure near maxima, 
and a substructure near one of the saddle-points, respectively.
For each case, we simulate~$10^3$ lens systems.
As expected from Figure~\ref{fig:oneNFW_eps_dist}, for each set,~\Rq values lies very
close to zero when there is no substructure present~(red points) which can also be seen 
from the corresponding histogram plot shown in Figure~\ref{fig:eNFWSub_mvir_hist}.
However, once we add a substructure near minima image, we can see from the corresponding
histogram plots that the average~\Rq values shift towards positive values in each 
set~(i.e., set-1, set-2, and set-3).
This shift increases as we go from set-1~(mean = 0.09) to set-3~(mean = 0.20).
This trend can be understood from the fact that in set-1, the source lies very close to
caustic, implying that the corresponding images all lie near to each other. 
Hence, a substructure which was placed near minima can also increase the magnification
of nearby saddle-point.
This also explains the tail in green histogram on the negative values.
On the other hand, in set-2 and set-3, images lie relatively far from each other in the
image plane implying that a substructure near minima will not affect the other other
images.
Here we stress that such a behaviour, substructure lying near minima
and boosting one of the saddle-points, is not common. 
Such an effect is only possible when the source lies near the tangential caustic so 
that minima and one of the saddle-point lies close to each other (for example, see image 
formation for red star in bottom panel of Figure~\ref{fig:ImageHU}).

Instead of minima, if we place a substructure near one of the saddle-point~(black points), 
we notice that for set-1,~2,~3, the average~\Rq value moves slightly~(but still consistent with zero) towards positive~(mean = 0.02), 
negative~(mean = -0.01), positive~(mean = 0.03) values, respectively.
For set-1, the change can be a result of either suppression of saddle-point or 
magnification of minima if the minima and saddle-point lie close to each other.
For set-2 and set-3, as the images lie relatively far from each other in the image plane,
the shift will be determined by the fraction of de-magnified area between the two
triangle-shaped caustic and the magnified area.
For set-1, when the substructure lies near saddle-point, we also observe a tail in~$R_{\rm hu}$
in the negative values in the corresponding histogram plot.
This can be explained by the above argument that sometimes the saddle-points images can
also be magnified by the presence of the substructure.

In the case of substructure near maxima~(yellow points), we observe that the~\Rq values 
moves towards the negative values (especially in set-1).
This can be understood as a result of over-focusing. 
When we add a substructure near an image, it increases the local surface mass density and, 
in turn, the local convergence, increasing the focusing of light rays. 
For maxima, this is equivalent to moving the image towards the centre of the lens where
the magnification is lower.
For set-2 and set-3, as the images are forming far from each other and critical curves,
implying that they are less magnified compared to set-1, and the deviations introduced
by the substructure are not strong enough to shift~\Rq sufficiently from zero.
In addition, we can see that the scatter around mean value~(i.e.,~$1\sigma$~values) are
larger in set-1 compared to set-2 and set-3.
This can be understood from the fact that the images in set-1 will be more magnified
than set-2 and set-3, implying that substructures which are relatively far from the images
can also lead to noticeable effects on image magnifications and flux ratios.
We also note that we add the substructure within the~5$\times$Einstein angle from the 
image implying that from the above analysis, we can only say that the substructure mass
range considered above is able to affect the average~$R_{\rm hu}$ values in a given HU 
image formation.
However, to estimate the properties of the underlying substructure, one needs to focus
on individual systems.

%%%%%%%%%%%%%%%%%%%%%%%%%%%%%%%%%%%%%%%%%%%%%%%%%%%%%%%%%%%%%%%%%%%%%%%%%%%%%%%%%%%%%%%%%%%
\section{\lowercase{e}NFW Lens + Many (Galaxy-scale) Substructures}
\label{sec:OneNFWMultipleSub}

In the case of real (observed) cluster lenses, we have a large number of galaxy scale
substructures with the number varying based on the mass of the cluster lens.
This implies that the above analysis that we have done for an isolated eNFW lens or 
an eNFW lens with one substructure will not be applicable to the actual cluster 
lenses unless the HU image formation is sufficiently isolated from cluster galaxies.
Hence, to get a better understanding of variation in~\Rq in actual cluster lenses, in
this section, we simulate cluster scales lenses with more than one substructure 
following simplistic assumptions.
We again focus on single halo cluster lenses where the main halo is represented by 
an eNFW lens with lens mass~${\rm M_{\rm vir}}\in[10^{14}, 10^{15}]~{\rm M_\odot}$, 
ellipticity $\epsilon = 0.3, 0.4, 0.5$ and concentration parameter~$c_{\rm vir}\in[1,20]$.
To populate the main halo with galaxy-scale perturbers, we assume the velocity 
dispersion~($v_{d}$), core radius~($\theta_{core}$), and cut-off radius~($\theta_{cut}$) 
of an~$L_*$ at the cluster redshift to be 104~km/s, 0.2~kpc, and 50~kpc, respectively.
Here we stress that~$(v_{d}, \theta_{core}, \theta_{cut})$ can be
different for different cluster lenses. For example,~\citet{2021MNRAS.508.1206J} 
derived higher values of these parameters for three different from detailed lens
modeling.
After that, we draw 100 galaxy-scale lenses from the mass 
range~$[10^9, 10^{13}]~{\rm M_\odot}$ with the probability 
distribution~$P(m)\propto(M_*/m)\exp[-m/M_*]$~\citep[e.g.,][]{1993ApJ...407L..49B} 
with~$M_*$ being the mass of~$L_*$ galaxy.
These galaxies were first randomly distributed in the cluster, and then we used the 
inverse of their distance to determine their position.
Doing so decreases the number of galaxies as we move outwards from the centre of the
cluster.
These galaxies have a total mass around~$\sim2\%$ of cluster halo mass.
Due to the presence of this additional mass in the cluster, typically, the HU will get
critical on a slightly lower redshift compared to the no-galaxy case.
We model these galaxies assuming a pseudo-isothermal elliptical mass 
distribution~\citep[PIEMD;][]{2007arXiv0710.5636E} profile.

Similar to the above, we again consider three different cases: 
(i) set-1: corresponding to source lying within 5~kpc of HU singularity,
(ii) set-2: corresponding to the source lying anywhere in the five-image region in the 
source plane but redshift is in the~$z_{\rm hu}\pm \Delta z$,
(iii) set-3: corresponding to generic five image formation.
The results are presented in Figure~\ref{fig:nfwMany_mvir_dist} for varying lens
mass while fixing~$c_{\rm vir} = 10$ and~$\epsilon=0.3, 0.4, 0.5$ for~$10^3$ simulated 
systems.
The corresponding histograms are shown in Figure~\ref{fig:nfwMany_mvir_hist}.
Interestingly, again we note that for a source lying near HU~(set-1), the average~$R_{\rm hu}$
value lies very close to zero for all three ellipticities, implying that most simulated 
systems still lead to~$R_{\rm hu}$ close to zero.
However, compared to the no-substructure case~($\sigma \approx 0.01$), the scatter around the 
mean value is significantly larger~($\sigma \approx 0.1$).
The large fraction of systems leading to~$R_{\rm hu} \approx 0$  implies that in most
cases, the chances of lying a galaxy-scale substructure near one of the images are low 
whereas the large scatter around zero implies the random distribution of the galaxy-scale 
substructure, which can shift~$R_{\rm hu}$ on either side of zero based on their position.

In set-2, we start to see the shifts in the average values for different ellipticities.
When~$\epsilon=0.3$, we see a slight shift in~\Rq towards negative values.
This can be understood from the fact that sources lying near the centre of the source 
plane will lead to de-magnified maxima images very near the lens centre, shifting the
\Rq towards negative values.
From Figure~\ref{fig:VaryHUeNFW}, increasing the ellipticity moves the~$z_{\rm hu}$ towards
larger redshifts.
However, doing so also pushed the saddle images towards the lens centre, decreasing their
magnification factors.
This shifts the overall~\Rq towards positive values.
We can see the effects in set-2, but they are more pronounced in generic five-image 
formation~(set-3).
Again we see that the scatter around the average~\Rq value is largest in the set-1 and
decrease down as we move to generic five image formation~(set-3), which can again be 
explained using the argument that images in set-1 are more magnified compared to set-3.
Results corresponding to variation in~$c_{\rm vir}$ instead of~$M_{\rm vir}$ are shown in
Figure~\ref{fig:nfwMany_cvir_dist} and~\ref{fig:nfwMany_cvir_hist} showing similar trends.

%%%%%%%%%%%%%%%%%%%%%%%%%%%%%%%%%%%%%%%%%%%%%%%%%%%%%%%%%%%%%%%%%%%%%%%%%%%%%%%%%%%%%%%%%%%
\section{Abell 1703}
\label{sec:a1703}

\begin{figure}
    \centering
    \includegraphics[scale=0.53]{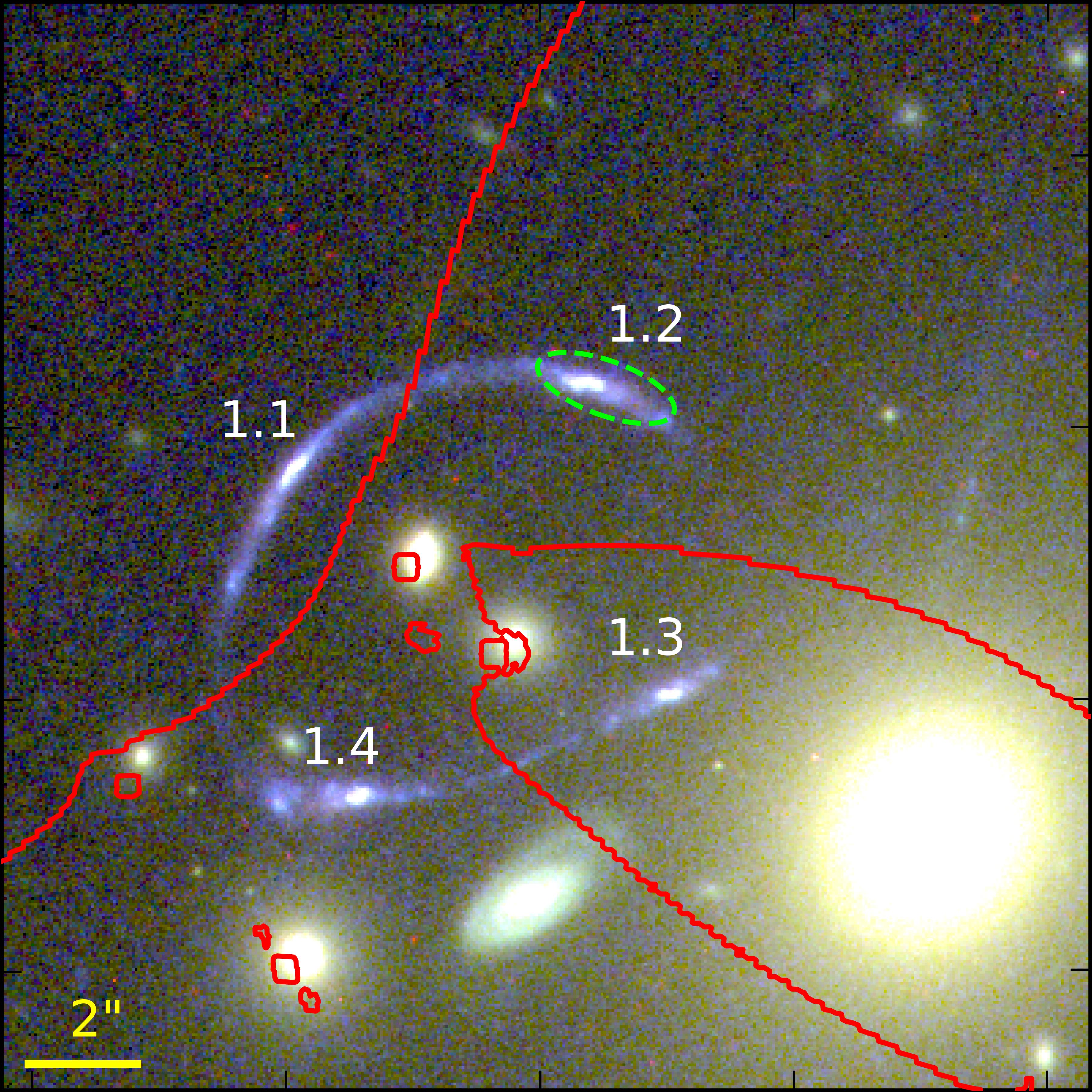}
    \caption{HU image formation in Abell 1703. The four lensed images of the source making 
    the HU image formation are marked in the same order as~\citet{2010MNRAS.408.1916Z}. Red 
    curves represent the critical curves at the source redshift, $z=0.889$. The green dashed
    ellipse around 1.2 marks the region in which we choose~$10^3$ random points as the lensed
    image position (see text for more details).}
    \label{fig:A1703_Image}
\end{figure}

Until very recently, only one HU image formation in the Abell~1703 galaxy cluster was 
observed~\citep[$z=0.28$;][]{2008A&A...489...23L}.   
The recent discovery of three HU image formations in a single galaxy cluster, 
RXJ0437+00~\citep{2023arXiv230309568L}, brings the total number of observed HU systems 
in clusters to four.
In this section, we analyse the~${R_{\rm hu}}$ for the HU image formation in Abell~1703 
cluster.
An RGB colour image of Abell~1703 is shown in Figure~\ref{fig:A1703_Image}. The images
part of the HU image formation are marked as 1.1, 1.2, 1.3, and 1.4.
The other image, 1.5, is on the other side of the cluster, and it is not part of the HU 
image formation.
We use the \textsc{light-trce-mass}~(\textsc{ltm}) lens model prenseted 
in~\citet{2010MNRAS.408.1916Z}.
The corresponding critical curves for a source at redshift $z=0.889$ are shown by red 
curves.
In our current work, we only use the best-fit lens model and do not include
the lens model led uncertainties on image position or magnification.

%%%%%%%%%%%%%%%%%%%%%%%%%%%%%%%%%%%%%%%%%%%%%%%%%%%%%%%%%%%%%%%%%%%%%%%%%%%%%%%%%%%%%%%%%%%

%%%%%%%%%%%%%%%%%%%%%%%%%%%%%%%%%%%%%%%%%%%%%%%%%%%%%%%%%%%%%%%%%%%%%%%%%%%%%%%%%%%%%%%%%%%
\begin{figure}
    \centering
    \includegraphics[scale=0.45]{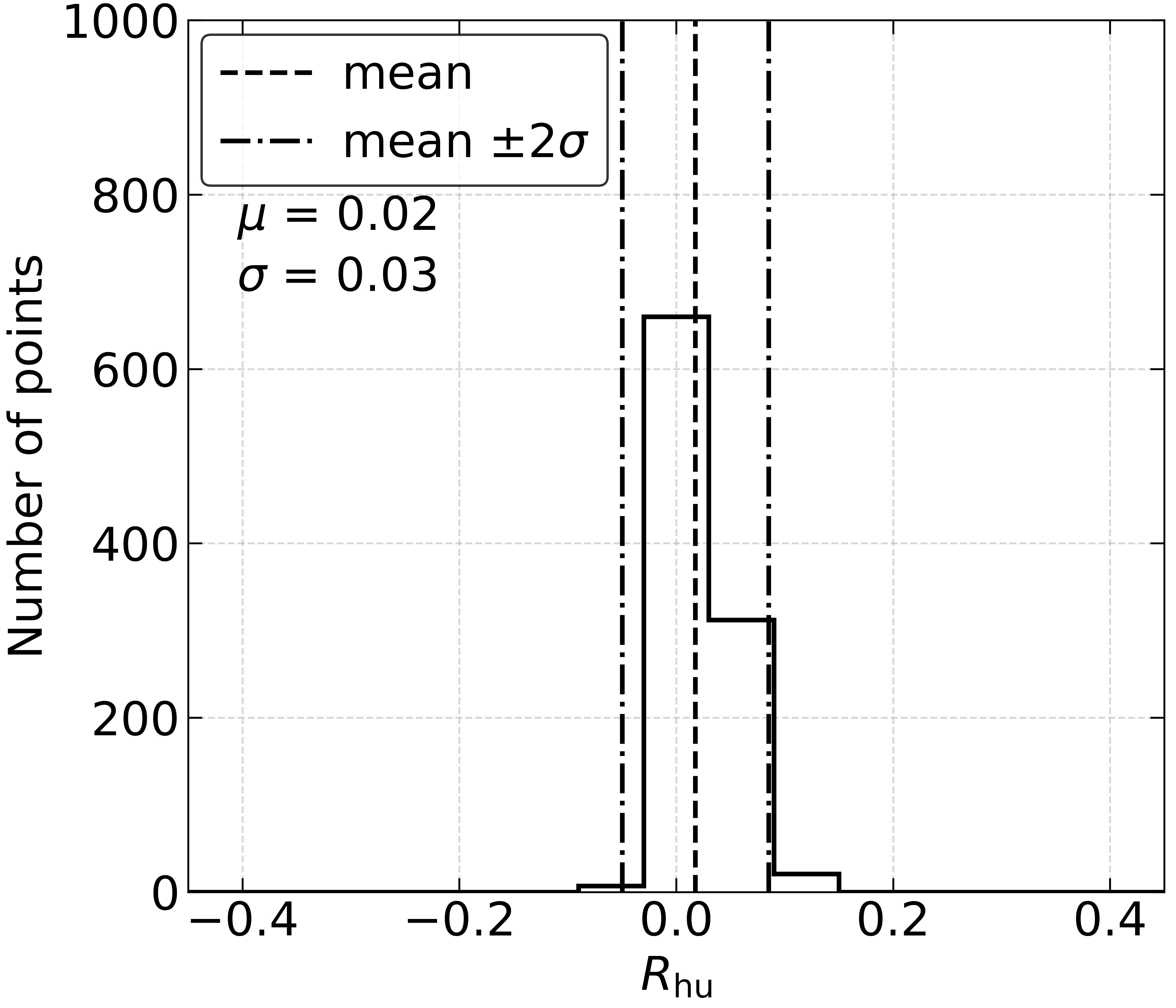}
    \caption{Histogram of~${R_{\rm hu}}$ values for 1000 random points in the source. The
    vertical dashed line represents the mean value of the histogram, whereas the 
    dashed-dotted lines cover the~$\pm 2 \sigma$ scatter around the mean. The numerical values
    of mean and~$1 \sigma$ are shown below the legend.}
    \label{fig:A1703_hist}
\end{figure}

%%%%%%%%%%%%%%%%%%%%%%%%%%%%%%%%%%%%%%%%%%%%%%%%%%%%%%%%%%%%%%%%%%%%%%%%%%%%%%%%%%%%%%%%%%%
\begin{figure*}
    \centering
    \includegraphics[scale=0.57]{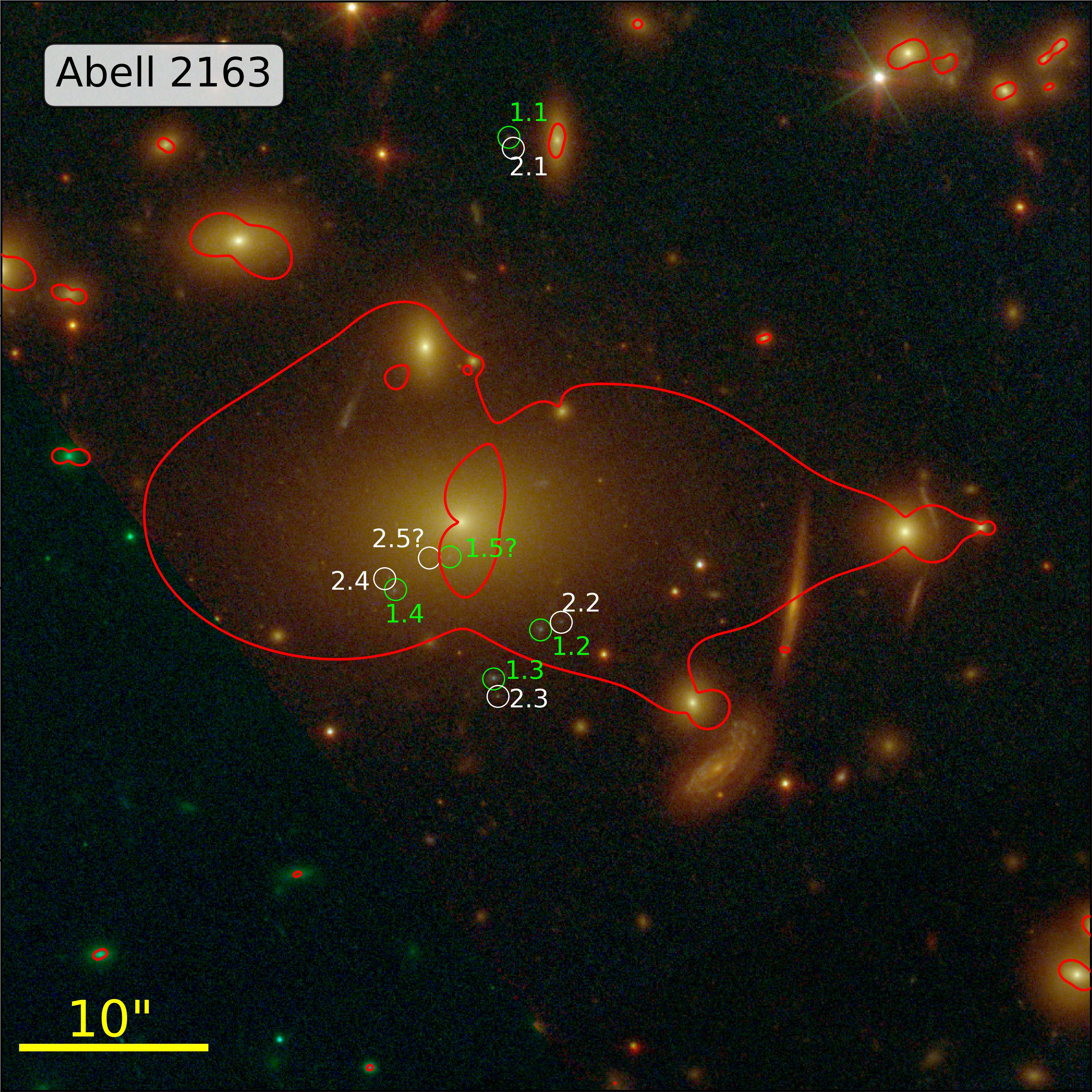}
    \includegraphics[scale=0.57]{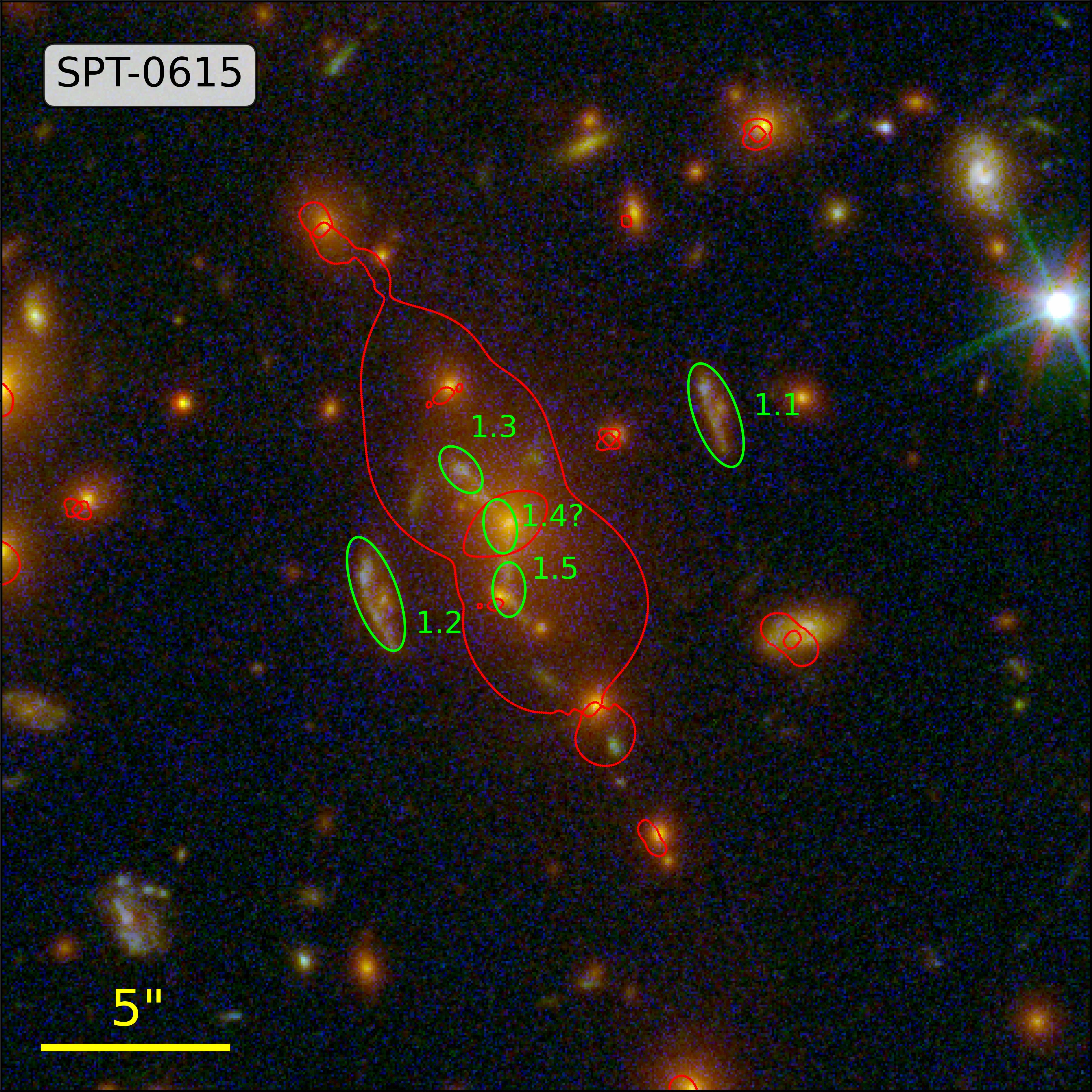}
    \caption{Three new HU candidate image formations in galaxy clusters. \textit{Left panel}. 
    Possible HU image formation in Abell~2163 galaxy cluster for two sources at 
    redshift~$\sim3$. The red curves show the critical curve for a source at redshift~3. 
    Images of HU candidates are marked by green and white circles.
    \textit{Right panel}. Possible Hu image formation in galaxy cluster SPT-CLJ0615–5746
    for a source at redshift~1.358, and the corresponding critical curves are shown by red
    curves. Images of HU candidates are marked by ellipses.}
    \label{fig:NewHU}
\end{figure*}

To estimate the flux ratio for this system, we randomly pick 1000 points inside the green
dashed ellipse on image 1.2 and shoot them back to the source plane.
Once we get the source positions, we again estimate the corresponding image positions.
Once we get the image positions, we remove the global minima (i.e., image~1.5)
and calculate the~$R_{\rm hu}$ from the remaining four images.
The histogram made using these 1000 points is shown in Figure~\ref{fig:A1703_hist}.
The vertical black dashed curve represents the mean of the histogram and dashed-dotted 
vertical lines cover the~$\pm2\sigma$ region around the mean.
The numerical values of mean~$(= 0.02)$ and~$1\sigma (=0.03)$ values are also shown in the plot.
From Figure~\ref{fig:A1703_hist}, we can see the histogram peaks very close to zero
and does not show a large scatter around it, which can also be seen from the~$\sigma$-value.
Comparing the mean value with mean values from Figure~\ref{fig:eNFWSub_mvir_hist}, we find
that the presence of a substructure in the mass range~$[10^7,10^{10}]~{\rm M_\odot}$ leads
to a larger flux ratio anomalies on average.
This implies, assuming the \textsc{ltm} model inferred~$R_{\rm hu}$ is correct, HU image 
formation in Abell~1703 can be used to detect substructures in the above mass range near 
any of the images if a substructure is present.

In principle, the flux ratio values and the corresponding scatter will also depend on the
lens modelling method. 
For example, in a non-parametric method such as \textsc{grale}~\citep{2006MNRAS.367.1209L}, 
the final best-fit model is averaged over many realisations smoothing out the galaxy-scale 
perturbers and diminishing their effects on flux ratio calculation\footnote{See~\citealt{2021MNRAS.503.2097M} and~\citealt{2021MNRAS.506.1526M}
for a discussion about effect of lens modeling method on the number of point singularities.}.
Hence, instead of \textsc{ltm} if we use \textsc{grale} lens model, we would expect even
smaller deviations from zero in~$R_{\rm hu}$ in Abell~1703 HU image formation.
On the other hand, if we consider a parametric model, then it might happen that the cluster
galaxies inside the ring formation in Figure~\ref{fig:A1703_Image} may have more
weights compared to~\textsc{ltm} model and might lead to a larger scatter in~$R_{\rm hu}$
values.
That said, a more detailed study is required to test the effect of lens
modeling technique on the~$R_{\rm hu}$.

%%%%%%%%%%%%%%%%%%%%%%%%%%%%%%%%%%%%%%%%%%%%%%%%%%%%%%%%%%%%%%%%%%%%%%%%%%%%%%%%%%%%%%%%%%%
\section{New HU candidates?}
\label{sec:NewHU}

In this section, we present three new candidates for HU image formations in two 
different galaxy clusters: Abell~2163 and~SPT-CLJ0615–5746.
Both of these clusters are observed by \textit{Hubble space telescope}~(HST) 
observations under the \textit{Reionization Lensing Cluster 
Survey}~\citep[RELICS;][]{2019ApJ...884...85C}\footnote{\url{https://archive.stsci.edu/hlsp/relics}} programme.
Colour images of both these clusters are shown in Figure~\ref{fig:NewHU}.
In the left panel, we show Abell~2163~($z=0.203$) leading to HU image formation for 
two sources lying at~$z\sim3$~\citep[e.g.,][]{2018ApJ...859..159C}.
Both sources are very compact and lie very close to each other and the corresponding
images are marked by green and white circles.
The red curves show the critical curves at redshift~$z=3$ for the \textsc{glafic} lens 
model.
In~\citet{2018ApJ...859..159C}, only the first three images of these sources are
mentioned as the lens images~(i.e., 1.1/2.1, 1.2/2.2, and 1.3/2.3).
However, the~\textsc{glafic} lens model predicts five images for each source with 
the fourth image~(1.4/2.4) being easily observed in the colour images. 
The lens model predicts the fifth image~(1.5/2.5) lying very close to the lens centre.
We do observe two clumps near the lens centre which are marked as~1.5/2.5.
However, we are not sure whether these belong to the lens system or if they are
some clumps in the lens itself.
Hence, we denote them with a question mark.
In the right panel of Figure~\ref{fig:NewHU}, we show the~SPT-CLJ0615–5746~($z=0.972$)
galaxy cluster with a source at redshift~$z=1.358$~\citep[e.g.,][]{2018ApJ...863..154P} 
leading to HU-like image formation.
We mark the images with green ellipses and red curves are critical curves for
a source at redshift~1.358.
We again observe that the predicted central maxima image~(i.e., 1.4?) lies very 
close to the cluster centre making it hard to detect.
However, from the critical curves, we can be sure about the formation of the
central image.

These two clusters present two different types of image formation possible near
the HU singularity.
However, further analysis is required, to ascertain whether these systems can be 
classified as HU image formations and their fitness in constraining the inner mass 
slope of these clusters, as well as detecting substructures near these image formations.

%%%%%%%%%%%%%%%%%%%%%%%%%%%%%%%%%%%%%%%%%%%%%%%%%%%%%%%%%%%%%%%%%%%%%%%%%%%%%%%%%%%%%%%%%%%
\section{Conclusions}
\label{sec:conclusions}

In this work, we have investigated the magnification relation for image formation 
near the HU singularity. 
Unlike fold and cusp, HU is an unstable/point singularity that only appears for 
specific arrangement(s) in a given strong lens system.
To better understand the conditions under which an HU is likely to appear, we have 
studied the effect of various parameters of an isolated eNFW lens on critical HU 
redshift.
In addition, to understand the behaviour of flux ratio in HU image systems, we 
studied the effect of various eNFW lens parameters on the~$R_{\rm hu}$ and compared 
it to the flux ratios in generic five-image formations.
After that, we analysed the effect of an individual substructure lying near one of 
the HU image on the flux ratio as well as effects of randomly distributed galaxy scale 
substructures in the lens.
We also calculate the~$R_{\rm hu}$ for the HU image formation in the Abell~1703 
cluster lens using the \textsc{ltm} lens model. 
Lastly, we present three possible HU image formation candidates in two different
RELICS cluster.
Based on the above analysis, our main findings are as follows:
\begin{itemize}
    \item eNFW lenses with small ellipticities values~($\epsilon<0.3$) or large 
    concentration parameter~($c_{\rm vir}>10$) are less likely to lead to the HU 
    image formation. This is because for such parameter values, critical HU 
    redshift~($z_{\rm hu}$) lies very close to the lens redshift, which 
    decreases the cross-section of the HU image formation.

    \item For eNFW lenses with ellipticities~$(\epsilon)\gtrsim0.3$, the 
    magnification relation~($R_{\rm hu}$) for HU image formation attains values very close 
    to zero with very small deviations~(standard deviation; $\sigma\lesssim0.02$) 
    and only shows a mild dependency on lens mass or concentration parameters.

    \item The existence of external shear has
     the capacity to modify the critical redshift for the HU 
    image~($z_{\rm hu}$) by affecting the distance from the lens centre where the 
    HU image is generated. Nevertheless, unless the external influences are very 
    strong, it is unlikely for them to cause a noteworthy deviation in $R_{\rm hu}$ 
    relative to the situation where no external shear is present.

    \item Presence of a substructure with mass~$[10^7,10^{10}]~{\rm M_\odot}$ 
    within 5~Einstein radii of one of the images, which are part of HU image 
    formation, can modify the~$R_{\rm hu}$ value depending on the type of image. 
    For example, a substructure near a minima image increases the corresponding 
    magnification, in turn, shifting~$R_{\rm hu}$ towards positive values. On the 
    other hand, if the substructure lies near the maxima image, then due to 
    over-focussing, the maxima image gets de-magnified, shifting~$R_{\rm hu}$ 
    towards negative values. A substructure present near one of the saddle-point
    de-magnifies it, shifting the~$R_{\rm hu}$ towards positive values. However,
    less probable, the substructure also has the capability to magnify the
    saddle-point moving the~$R_{\rm hu}$ towards negative values.
    
    \item Comparison of~$R_{\rm hu}$ with~\Rq (in set-2 and set-3) in the presence
    of substructure reveals that~\Rq shows smaller deviations from zero compared 
    to~$R_{\rm hu}$ and less scatter around zero. This is due to a decrease in
    image magnification for~set-2 and set-3 implying a large deviation from
    HU image formation might not be useful for substructure analysis.
    
    \item A population of substructures within the main halo (equivalent to the 
    cluster galaxies) on average does not introduce a shift in the~$R_{\rm hu}$ 
    from zero, but it does increase the scatter around the average value. This 
    implies that the overall chances of lying a galaxy-scale lens near any of 
    the HU images such that it can significantly affect the~$R_{\rm hu}$ is small.
    This also implies that we can expect a small deviation from zero in a large number
    of HU image formations which can help us to detect substructures in the lens.
    We note that this inference is based on the average behaviour of the~$R_{\rm hu}$.
    However, for robust inferences about substructure properties near the observed HU
    image formations, we need to focus on individual systems.

    \item For a population of substructures within the main halo (equivalent to 
    the cluster galaxies), a source lying anywhere in the five-image region with 
    redshift in the range~$z_{\rm hu} \pm \Delta z$~(set-2), the average~\Rq 
    slightly shifts towards positive or negative values depending on the lens 
    ellipticity. This effect is even more noticeable in the generic five-image 
    formations~(set-3). In addition, as we go from set-1 to set-3, the standard
    deviation around the mean decrease due to the decrease in image magnification.

    \item Using \textsc{ltm} lens model for Abell~1703, we find that the average 
    value of~$R_{\rm hu}$ is 0.02 (which is very close to zero) for the 
    corresponding HU image formation with a standard deviation~($\sigma$) of~0.03. 
    These findings suggest that the likelihood of detecting substructures close 
    to the HU image formation is high, especially when compared to the eNFW$+$~one 
    substructure scenario.

    \item We present three new candidates of HU image formations in RELICS clusters,
    two in Abell~2163 and one in~SPT-CLJ0615–5746. For all of these systems, the
    central maxima image of the HU image formation is expected to lie very close 
    to the centre of the cluster. Further study of these systems is required to
    determine whether these systems are actually HU images formation and if are
    suitable for finding substructures near them.
\end{itemize}

So far, all of the HU image formations are detected in the HST observations 
of galaxy clusters.
However, to detect a small-scale substructure near a HU image formation, one 
also needs to detect the small-scale clumps in the source itself so that we can 
calculate the corresponding magnification relation. 
The size of the source galaxy is much larger than the Einstein radius of the 
substructure diminishing its effects.
Thanks to the larger photon collecting area compared to HST, 
JWST~\citep{2006SSRv..123..485G} can bring forth a large number of these small-scale 
clumps of light present in the lensed 
sources~\citep[e.g.,][]{2022arXiv221014123H, 2023MNRAS.520.2180C}. 
Hence, once a sufficiently large sample of HU image formations is identified in the 
sky, observations with JWST can lead to the identification of a large number of 
small-scale clumps in these HU image formations, increasing our chances of 
detecting substructures lensing.
Observation of clusters lenses with JWST is also beneficial in making more 
accurate lens models by identifying new lensed 
systems~\citep[e.g.,][]{2022ApJ...938L...6P, 2023A&A...672A...3D, 2023ApJ...944L...6M} 
and allowing us to better differentiate the effect of the overall cluster lens 
and possible substructure~(near HU image) itself.

In our current work, we have mainly focused on the HU image formation in isolated 
eNFW lenses.
However, merging clusters can further boost the chances of detecting HU systems 
as each eNFW component of such a cluster is able to form a pair of HU 
singularities at different redshifts.
But the presence of a second component can also decrease the critical HU redshift 
which can actually decrease the HU cross-section instead of increasing it.

We have not discussed the possibility of detecting HU image formation in the 
galaxy scale lenses.
Since we expect more galaxy-galaxy lens systems compared to cluster lenses, just 
in sheer number, there are more HU singularities in galaxy lenses.
However, as the galaxy lenses are less massive and are (on average) less elliptical 
compared to cluster lenses, there is a risk of HU images being lost in the light 
of the lens.  
One might be able to overcome this problem with observations in sub-millimetre or 
radio bands which are not expected to be affected by the visible light coming from 
the lens itself.
That said, the above inferences for galaxy-scale lenses are based on oversimplified 
assumptions, and we are conducting more comprehensive analyses to determine if 
galaxy lenses can lead us to HU formations or not.

%%%%%%%%%%%%%%%%%%%%%%%%%%%%%%%%%%%%%%%%%%%%%%%%%%%%%%%%%%%%%%%%%%%%%%%%%%%%%%%%%%%%%%%%%%%
\section{Acknowledgements}
The authors are very grateful to Adi Zitrin for providing the data products related to 
Abell~1703.
Authors thank the anonymous referee for constructuve comments.
AKM acknowledges support by grant 2020750 from the United States-Israel Bi-national Science 
Foundation (BSF) and grant 2109066 from the United States National Science Foundation (NSF) 
and by the Ministry of Science $\&$ Technology, Israel.
This research has made use of NASA's Astrophysics Data System Bibliographic Services.
\\
\\
\textit{Software:}
\textsc{python}~(\url{https://www.python.org/}),
\textsc{astropy}~\citep{2018AJ....156..123A},
\textsc{matplotlib}~\citep{2007CSE.....9...90H},
\textsc{numpy}~\citep{2020Natur.585..357H},
\textsc{scipy}~\citep{2020SciPy-NMeth},
\textsc{shapely}~\citep{gillies_sean_2022_7428463}.

%%%%%%%%%%%%%%%%%%%%%%%%%%%%%%%%%%%%%%%%%%%%%%%%%%%%%%%%%%%%%%%%%%%%%%%%%%%%%%%%%%%%%%%%%%%
\section{Data Availability}
All the simulated data used in this work can be easily reproduced following the methods 
discussed in the text.
The colour image of Abell~1703 and the corresponding \textsc{ltm} lens model can be made 
available from Adi Zitrin on a reasonable request.

%%%%%%%%%%%%%%%%%%%%%%%%%%%%%%%%%%%%%%%%%%%%%%%%%%%%%%%%%%%%%%%%%%%%%%%%%%%%%%%%%%%%%%%%%%%
\bibliographystyle{mnras}
\bibliography{reference}

%%%%%%%%%%%%%%%%%%%%%%%%%%%%%%%%%%%%%%%%%%%%%%%%%%%%%%%%%%%%%%%%%%%%%%%%%%%%%%%%%%%%%%%%%%%
\appendix

%%%%%%%%%%%%%%%%%%%%%%%%%%%%%%%%%%%%%%%%%%%%%%%%%%%%%%%%%%%%%%%%%%%%%%%%%%%%%%%%%%%%%%%%%%%
\section{NFW Lens}
\label{app:LensNFW}

The virial mass of a lensing halo at redshift $z$ with virial radius $R_{\rm vir}$
is given as,
\begin{equation}
    M_{\rm vir} = \frac{4\pi}{3}R_{\rm vir}^3 \: 200 \: \rho_{\rm cr}(z),
    \label{eq:VirMass}
\end{equation}
where $\rho_{\rm cr}(z)$ is the critical density of the Universe at the redshift $z$.
The three-dimensional mass distribution of such a halo assuming an NFW profile 
is given as
\begin{equation}
    \rho(r) = \frac{\rho_s}{\left(r/r_s\right)\left(1+r/r_s\right)^2},
    \label{eq:Den3D}
\end{equation}
where $r_s$ are scale radius of the halo defining the concentration parameter
as $c_{\rm vir}\equiv R_{\rm vir}/r_s$ and $\rho_s$ is the density at the scale 
radius,
\begin{equation}
    \rho_s = \frac{200}{3} \rho_{\rm cr} \frac{c^3}{\ln(1+c)-c/(1+c)}.
    \label{eq:ScaleRadius}
\end{equation}
The projected gravitational lensing potential corresponding to an NFW halo is given
as (in angular units),
\begin{equation}
    \Psi(\theta) = 4\kappa_{\rm s} g(\theta),
    \label{eq:PotOneNFW}
\end{equation}
where
\begin{equation}
    g(\theta) = \frac{1}{2} \ln^2\frac{\theta}{2} + 
    \begin{cases}
        \quad \: 2\:{\rm atan^2\sqrt{\frac{\theta-1}{\theta+1}}}, &  \theta > 1\\
        -2\:{\rm atanh^2\sqrt{\frac{1-\theta}{1+\theta}}}, &  \theta < 1\\
        \qquad \qquad \qquad 0, & \theta=1
    \end{cases}
    \label{eq:gFunc}
\end{equation}
with $\theta=\sqrt{\theta_1^2+\theta_2^2/q^2}$ and $q$ is the axis ratio.
We define the lens ellipticity as $\epsilon\equiv1-q$.

%%%%%%%%%%%%%%%%%%%%%%%%%%%%%%%%%%%%%%%%%%%%%%%%%%%%%%%%%%%%%%%%%%%%%%%%%%%%%%%%%%%%%%%%%%%
\section{\emph{R-\lowercase{quantity}} \lowercase{vs.} ($M_{\rm \lowercase{vir}}, \lowercase{c}_{\rm \lowercase{vir}}$) \lowercase{for} \lowercase{e}NFW lens}
\label{app:LensNFW_MCvirDist}

%%%%%%%%%%%%%%%%%%%%%%%%%%%%%%%%%%%%%%%%%%%%%%%%%%%%%%%%%%%%%%%%%%%%%%%%%%%%%%%%%%%%%%%%%%%
\begin{figure*}
    \centering
    \includegraphics[scale=0.58]{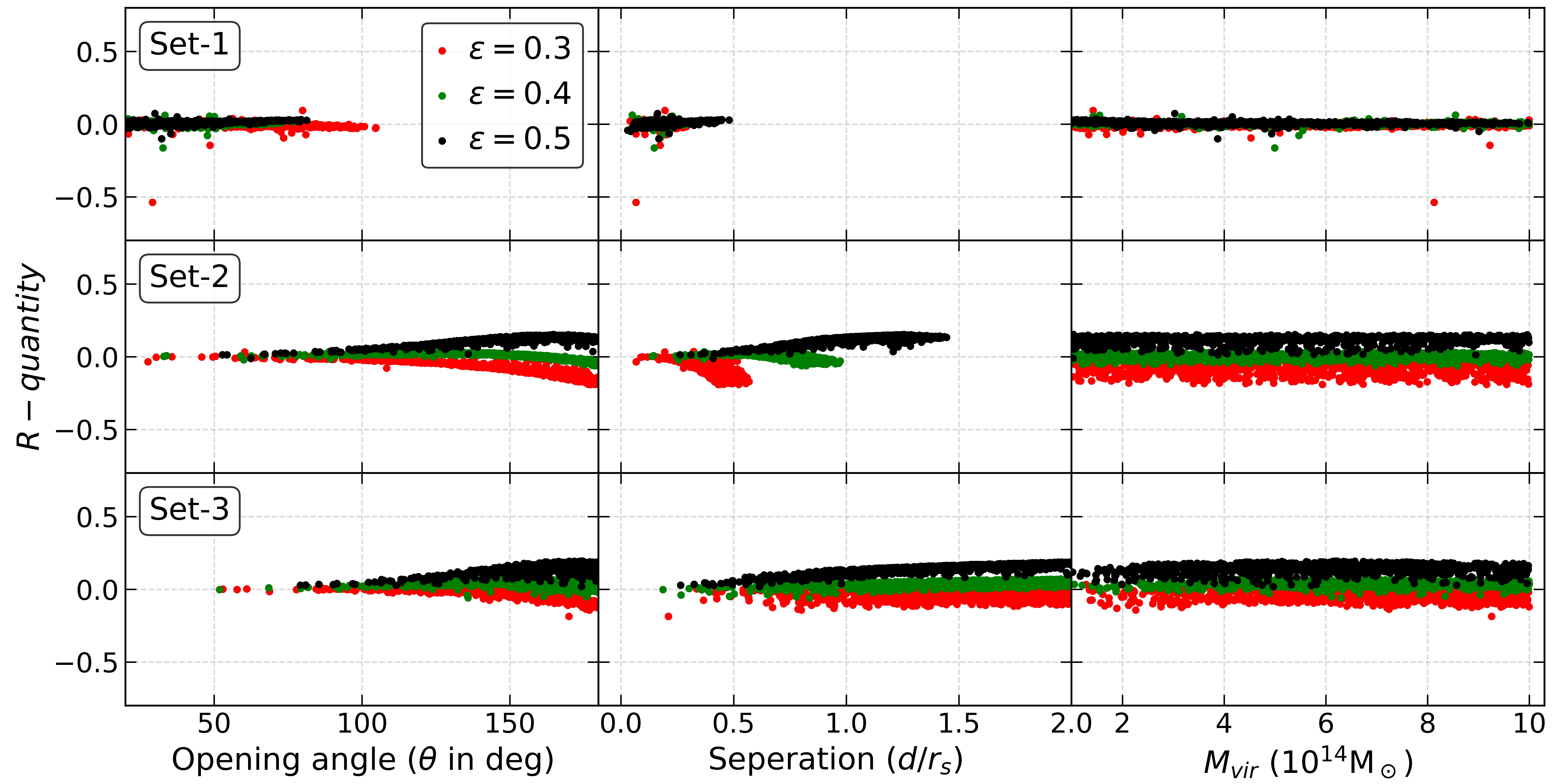}
    \caption{\Rq as a functions of lens mass~($M_{\rm vir}$) for an eNFW lens with 
    concentration parameter~($c_{\rm vir}$) fixed to~10. In each panel, red, green, 
    and black points correspond to~$\epsilon=0.3, 0.4,~{\rm and}~0.5$, respectively.
    The top, middle, and bottom rows represents the set-1~(where \Rq is essential 
    represents~$R_{\rm hu}$), set-2, and set-3, respectively. In the left, middle, 
    and right columns, the \emph{R-quantity} is plotted as a function of image opening 
    angle~($\theta$), maximum image separation~($d$; in units of~$r_s$), and lens 
    mass~($M_{\rm vir}$), respectively.}
    \label{fig:oneNFW_mvir_dist}
\end{figure*}

%%%%%%%%%%%%%%%%%%%%%%%%%%%%%%%%%%%%%%%%%%%%%%%%%%%%%%%%%%%%%%%%%%%%%%%%%%%%%%%%%%%%%%%%%%%
\begin{figure*}
    \centering
    \vspace{1cm}
    \includegraphics[scale=0.58]{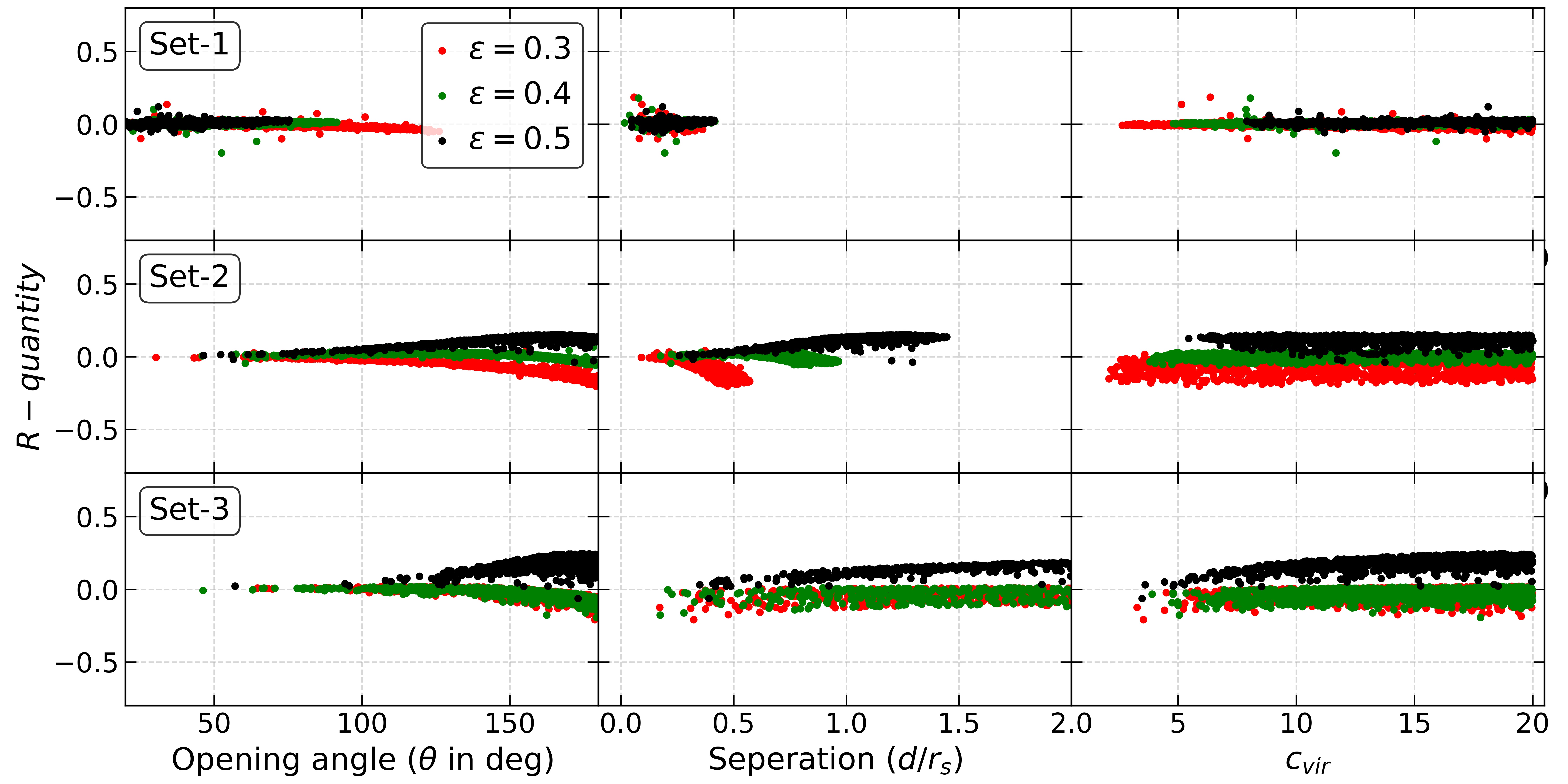}
    \caption{\Rq as a functions of concentration parameter~($c_{\rm vir}$) for an eNFW 
    lens with lens mass~($M_{\rm vir}$) fixed to~$5\times10^{14}~{\rm M_\odot}$. In each 
    panel, red, green, and black points correspond to~$\epsilon=0.3, 0.4,~{\rm and}~0.5$, 
    respectively.
    The top, middle, and bottom rows represent the set-1~(where \Rq is essential 
    represents~$R_{\rm hu}$), set-2, and set-3, respectively. In the left, middle, 
    and right columns, the \emph{R-quantity} is plotted as a function of image opening 
    angle~($\theta$), maximum image separation~($d$; in units of~$r_s$), and lens 
    mass~($c_{\rm vir}$), respectively. In the bottom panel, we can observe 
    a sudden jump in~\Rq values for~$\epsilon=0.3/0.4$ and~$\epsilon=0.5$. It can again be 
    understood from the fact that for large~$\epsilon$ values, the saddle-point images will 
    form near the lens centre and get de-magnified, as discussed in Section~\ref{ssec:rhu_enfw_eps}.}
    \label{fig:oneNFW_cvir_dist}
\end{figure*}

%%%%%%%%%%%%%%%%%%%%%%%%%%%%%%%%%%%%%%%%%%%%%%%%%%%%%%%%%%%%%%%%%%%%%%%%%%%%%%%%%%%%%%%%%%%
\section{\emph{R-\lowercase{quantity}} \lowercase{vs.} $\lowercase{c}_{\rm \lowercase{vir}}$ \lowercase{for} \lowercase{e}NFW \lowercase{+} one substructure lens}
\label{app:LensNFWSS_CvirDist}

%%%%%%%%%%%%%%%%%%%%%%%%%%%%%%%%%%%%%%%%%%%%%%%%%%%%%%%%%%%%%%%%%%%%%%%%%%%%%%%%%%%%%%%%%%%
\begin{figure*}
    \centering
    \includegraphics[scale=0.58]{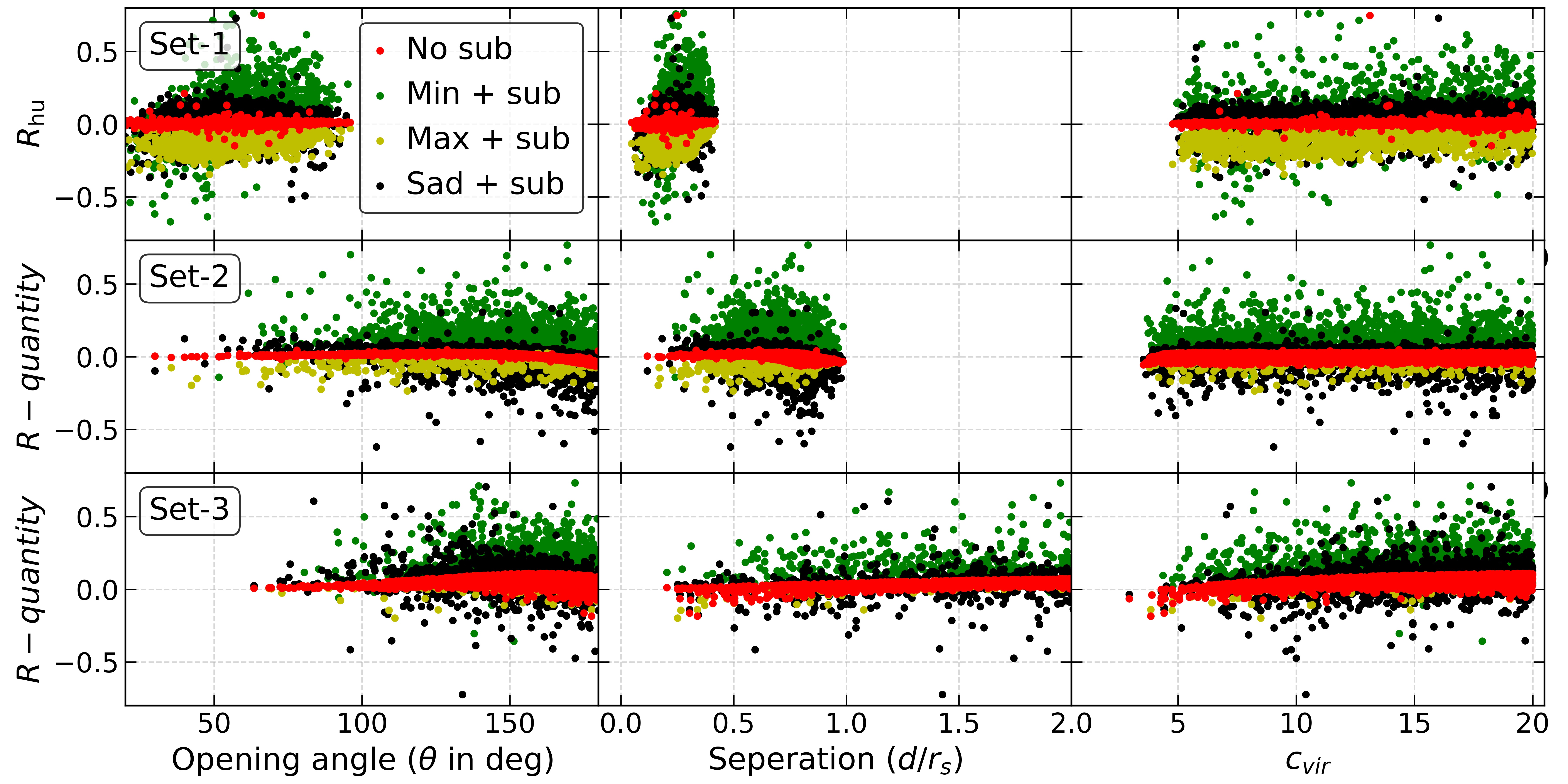}
    \caption{\Rq as a functions of main lens concentration parameter~($c_{\rm vir}$) for 
    an eNFW lens~$+$~one substructure with main lens mass~($M_{\rm vir}$) fixed 
    to~$5\times10^{14}~{\rm M_\odot}$. In each panel, red, green, yellow, and black points 
    correspond to no substructure, substructure near minima, substructure near maxima, and substructure near 
    one of the saddle-points, respectively.
    The top, middle, and bottom rows represent the set-1~(where \Rq is essential 
    represents~$R_{\rm hu}$), set-2, and set-3, respectively. In the left, middle, 
    and right columns, the \emph{R-quantity} is plotted as a function of image opening 
    angle~($\theta$), maximum image separation~($d$; in units of~$r_s$), and lens 
    mass~($c_{\rm vir}$), respectively.}
    \label{fig:oneNFWSS_cvir_dist}
\end{figure*}

%%%%%%%%%%%%%%%%%%%%%%%%%%%%%%%%%%%%%%%%%%%%%%%%%%%%%%%%%%%%%%%%%%%%%%%%%%%%%%%%%%%%%%%%%%%
\begin{figure*}
    \centering
    \vspace{1cm}
    \includegraphics[scale=0.58]{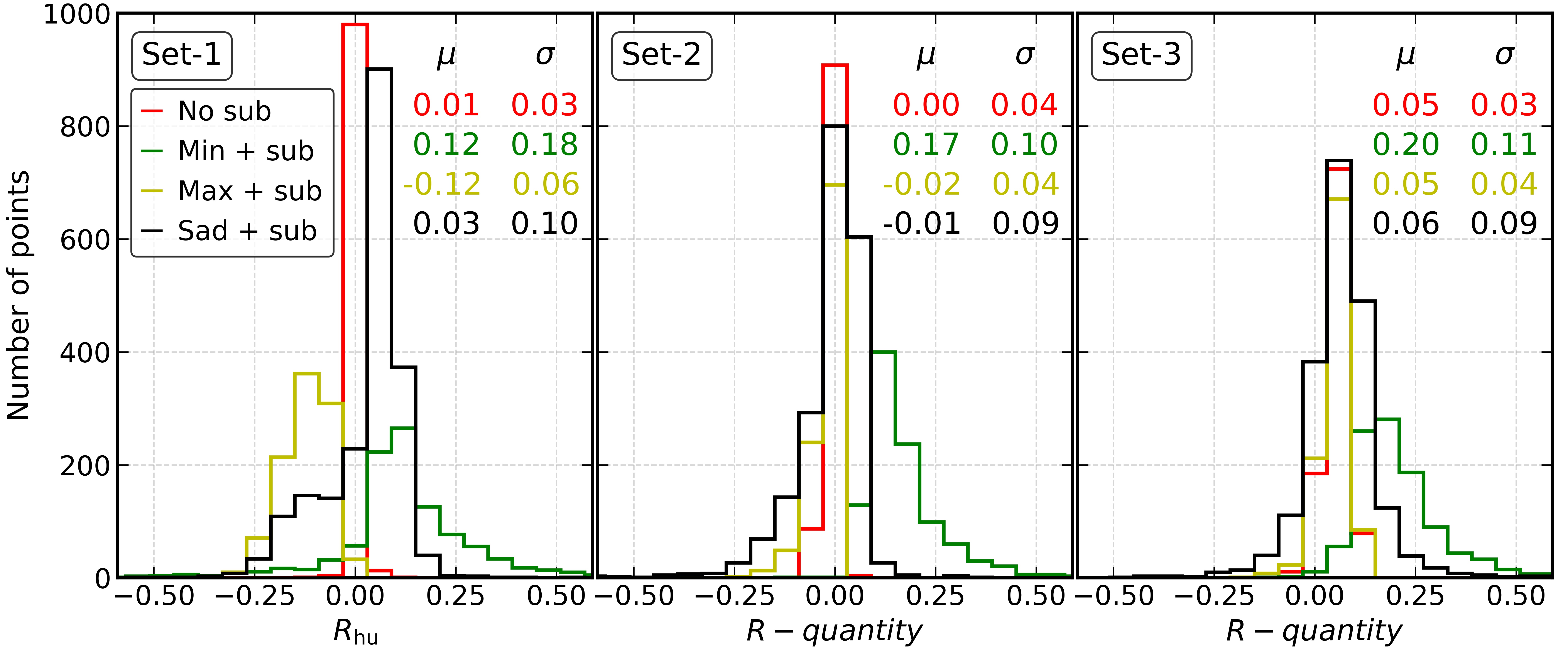}
    \caption{\Rq histogram plot for systems shown in Figure~\ref{fig:oneNFW_cvir_dist}.
    The left, middle, and right panels are corresponding to set-1, set-3, and set-3, 
    respectively. Similar to Figure~\ref{fig:oneNFW_cvir_dist}, the red, green, yellow,
    and black histograms are corresponding to no substructure, substructure near minima, substructure
    near maxima, and substructure near one of the saddle-points, respectively. For each 
    histogram, the mean~($\mu$) and standard deviation~($\sigma$) are shown in the upper 
    right part of each panel.}
    \label{fig:oneNFWSS_cvir_hist}
\end{figure*}

%%%%%%%%%%%%%%%%%%%%%%%%%%%%%%%%%%%%%%%%%%%%%%%%%%%%%%%%%%%%%%%%%%%%%%%%%%%%%%%%%%%%%%%%%%%
\section{\emph{R-\lowercase{quantity}} \lowercase{vs.} $\lowercase{c}_{\rm \lowercase{vir}}$ \lowercase{for} \lowercase{e}NFW \lowercase{+} many substructures lens}
\label{app:LensNFWMany_CvirDist}

%%%%%%%%%%%%%%%%%%%%%%%%%%%%%%%%%%%%%%%%%%%%%%%%%%%%%%%%%%%%%%%%%%%%%%%%%%%%%%%%%%%%%%%%%%%
\begin{figure*}
    \centering
    \includegraphics[scale=0.58]{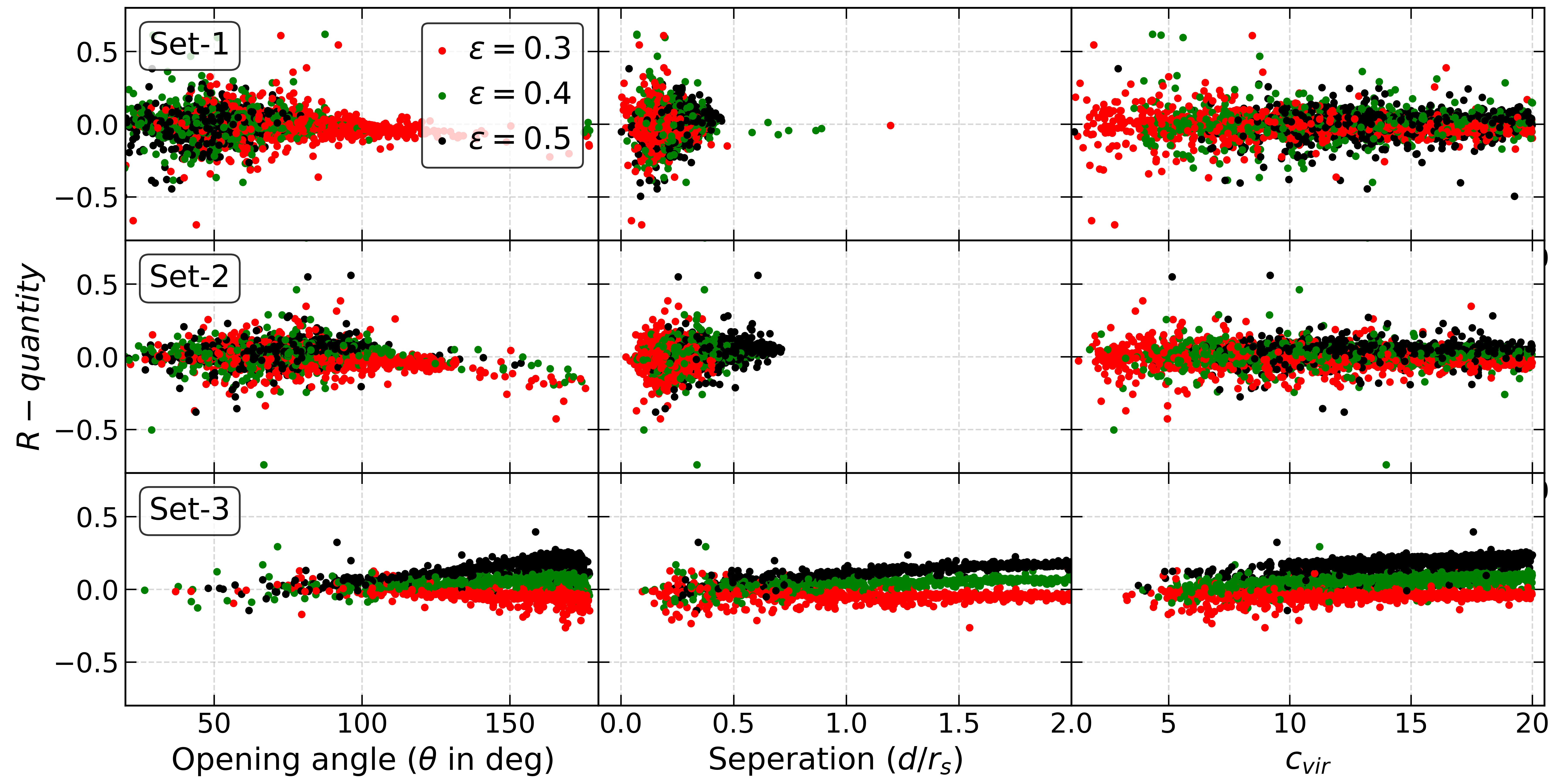}
    \caption{\Rq as a functions of main lens concentration parameter~($c_{\rm vir}$) for 
    an eNFW lens~$+$~many substructure with main lens mass~($M_{\rm vir}$) fixed 
    to~$5\times10^{14}~{\rm M_\odot}$. In each panel, red, green, and black points 
    correspond to~$\epsilon=0.3, 0.4,~{\rm and}~0.5$, respectively.
    The top, middle, and bottom rows represent the set-1~(where \Rq is essential 
    represents~$R_{\rm hu}$), set-2, and set-3, respectively. In the left, middle, 
    and right columns, the \emph{R-quantity} is plotted as a function of image opening 
    angle~($\theta$), maximum image separation~($d$; in units of~$r_s$), and lens 
    mass~($c_{\rm vir}$), respectively.}
    \label{fig:nfwMany_cvir_dist}
\end{figure*}

%%%%%%%%%%%%%%%%%%%%%%%%%%%%%%%%%%%%%%%%%%%%%%%%%%%%%%%%%%%%%%%%%%%%%%%%%%%%%%%%%%%%%%%%%%%
\begin{figure*}
    \centering
    \vspace{1cm}
    \includegraphics[scale=0.58]{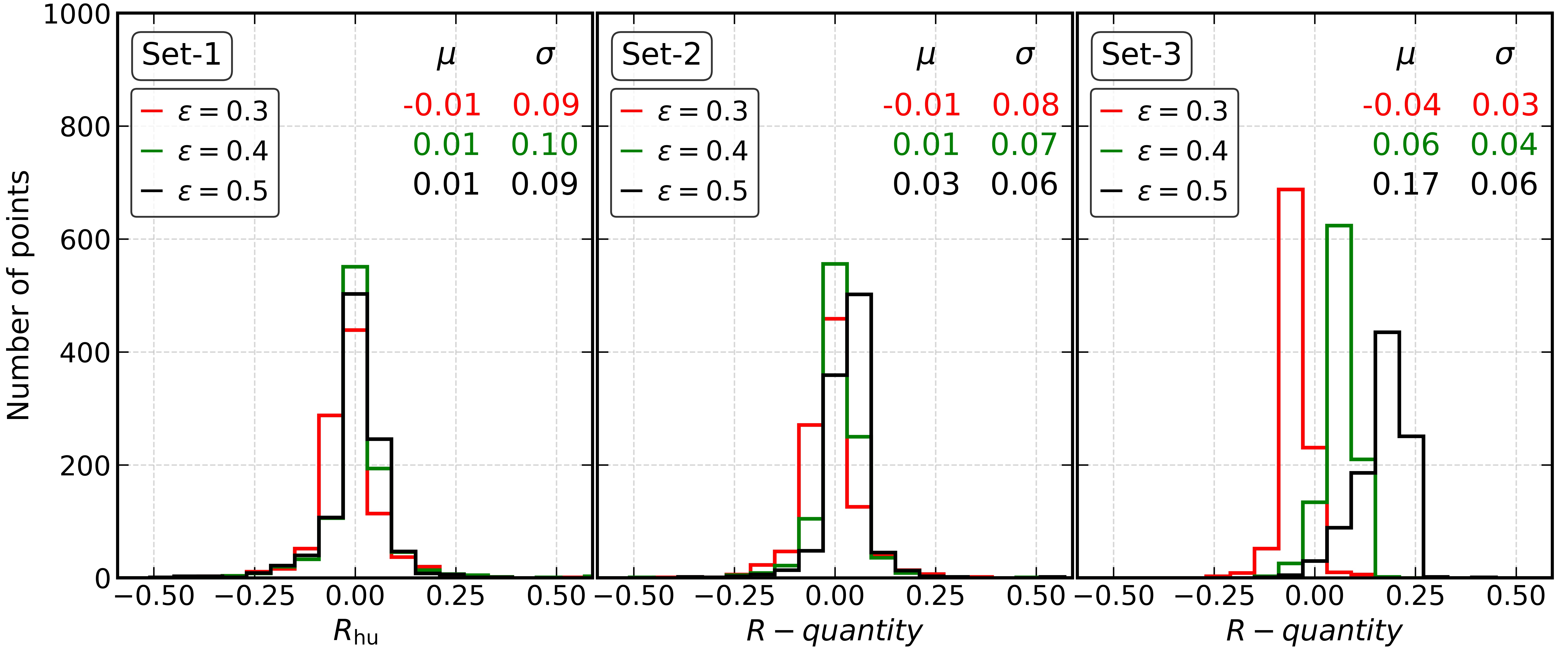}
    \caption{\Rq histogram plot for systems shown in Figure~\ref{fig:nfwMany_cvir_dist}.
    The left, middle, and right panels are corresponding to set-1, set-3, and set-3, 
    respectively. Similar to Figure~\ref{fig:nfwMany_cvir_dist}, the red, green, and 
    black histograms are corresponding to~$\epsilon=0.3, 0.4,~{\rm and}~0.5$, respectively. 
    For each histogram, the mean~($\mu$) and standard deviation~($\sigma$) are shown in 
    the upper right part of each panel.}
    \label{fig:nfwMany_cvir_hist}
\end{figure*}

%%%%%%%%%%%%%%%%%%%%%%%%%%%%%%%%%%%%%%%%%%%%%%%%%%%%%%%%%%%%%%%%%%%%%%%%%%%%%%%%%%%%%%%%%%%
% Don't change these lines
\bsp	% typesetting comment
\label{lastpage}
\end{document}